\documentclass[fleqn,10pt]{wlscirep}

\usepackage{float}
\usepackage{framed}
\usepackage{soul}
\usepackage{bm}
\usepackage{physics}
\usepackage{amsfonts}
\usepackage[utf8]{inputenc}
\usepackage[T1]{fontenc}
\usepackage{mathpazo} % get Palatino math  
\usepackage{amsmath}
\usepackage{textalpha}
\usepackage{newtxtext,newtxmath} % times math fonts (see https://tex.stackexchange.com/a/308148)

\DeclareUnicodeCharacter{FF08}{(}
\renewcommand{\vec}[1]{{\ensuremath{\bm{\mathrm{#1}}}}}

\newcommand{\rev}[1]{\textcolor{black}{#1}}
\newcommand{\note}[1]{\textcolor{black}{#1}}

\title{Dynamical stability by spin transfer\\ \rev{in nearly isotropic magnets}}

\author[1,2,3,4,5,*]{Hidekazu Kurebayashi}
\author[6]{Joseph Barker}
\author[4]{Takumi Yamazaki}
\author[4]{Varun K. Kushwaha}
\author[7]{Kilian D. Stenning}
\author[1]{Harry Youel}
\author[4]{\rev{Xueyao Hou}}
\author[3,4]{\rev{Troy Dion}}
\author[1]{\rev{Daniel Prestwood}}
\author[3,4,5,8]{Gerrit E. W. Bauer}
\author[9,*]{Kei Yamamoto}
\author[3,4,*]{Takeshi Seki}
\affil[1]{London Centre for Nanotechnology, University College London,  17-19 Gordon Street, London, WC1H 0AH, United Kingdom}
\affil[2]{Department of Electronic and Electrical Engineering, University College London, Roberts Building, London, WC1E 7JE, United Kingdom}
\affil[3]{Center for Science and Innovation in Spintronics, Tohoku University, 2-1-1, Katahira, Sendai, 980-8577 Japan}
\affil[4]{Institute for Materials Research, Tohoku University, 2-1-1, Katahira, Sendai, 980-8577 Japan}
\affil[5]{WPI Advanced Institute for Materials Research, Tohoku University, 2-1-1, Katahira, Sendai 980-8577, Japan}
\affil[6]{School of Physics and Astronomy, University of Leeds, Leeds LS2 9JT, United Kingdom}
\affil[7]{Blackett Laboratory, Imperial College London, London SW7 2AZ, United Kingdom}
\affil[8]{Kavli Institute for Theoretical Sciences, University of the Chinese Academy of Sciences, Beijing 100190, China}
\affil[9]{Advanced Science Research Center, Japan Atomic Energy Agency, 2-4 Shirakata, Tokai 319-1195, Japan}

\affil[*]{Emails: h.kurebayashi@ucl.ac.uk, yamamoto.kei@jaea.go.jp and takeshi.seki@tohoku.ac.jp}

\begin{abstract}
Spin transfer torques (STTs) control magnetisation by electric currents, enabling a range of nano-scale spintronic applications~\cite{RALPH_JMMM2008,Apalkov_IEEE2016,Hirohata_JMMM2020}. They can destabilise the equilibrium magnetisation state by counteracting magnetic relaxation. \rev{Here, we maximise the STT effect through a dedicated growth–annealing protocol for CoFeB thin films, such that magnetic anisotropies originating from the interface and shape almost cancel each other. The nearly isotropic magnets enable low-current dynamical stabilisation of the magnetisation in the direction opposite to an applied magnetic field, thereby realising a spintronic analogue of the Kapitza pendulum~\cite{Kapitza1951}. In an intermediate current regime, the STT drives large magnetisation vector fluctuations that cover the entire Bloch sphere.} The continuous variable associated with the stochastic magnetisation \rev{direction} may serve as a resource for probabilistic computing and neuromorphic hardware. Our results establish isotropic magnets as a platform to study as-yet-uncharted, far-from-equilibrium spin dynamics including anti-magnonics~\cite{Harms_AIPAdv2024}, with promising implications for unconventional computing paradigms~\cite{Chowdhury_IEEE2023}.
\end{abstract}

\begin{document}

\flushbottom
\maketitle

\section*{Introduction}

\begin{figure}[t]
    \centering
    \includegraphics[width=1.0\linewidth]{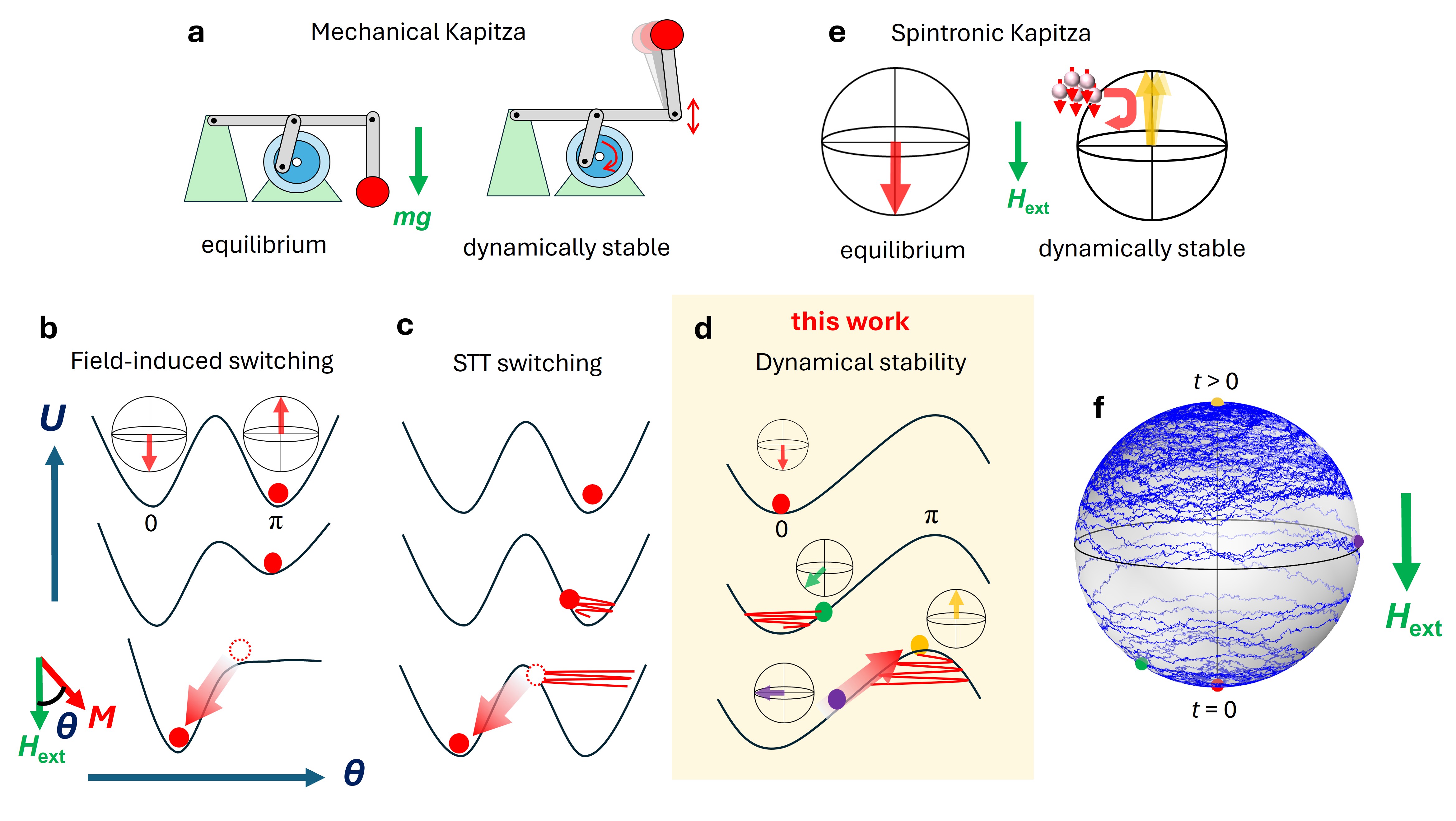}
  \caption{\textbf{Dynamical stability of the spintronic Kapitza pendulum.} \textbf{a}, Schematics of the mechanical Kapitza pendulum with and without an external drive. $mg$ represents the gravitational force on the bob. \textbf{b}-\textbf{d}, Schematics of different magnetisation switching and stability processes illustrated within the angular dependence of magnetic energy ($U$) (\textbf{b} field-induced switching, \textbf{c} STT switching, \textbf{d} dynamical stability). $\theta$ is defined by the relative angle between the magnetic field ($\bm{H}_{\rm ext}$) and moment ($\bm{M}$). See the main text for the mechanism of each process. \textbf{e}, Analogy between the Kapitza pendulum and dynamical stabilisation by spin transfer in an isotropic magnet. The magnetic field provides a minimum and maximum of potential energy as the gravitational field does, and the STT plays the role of dynamical driving that controls their stability. Note that the arrows piercing through the conduction electrons (solid spheres) represent their spin pointing opposite to their  magnetic moment. \textbf{f}, A typical solution of the stochastic LLG equation for a macrospin in an isotropic magnet. The initial condition is set at the south pole (red dot) and due to the anti-damping  STT exceeding the Gilbert damping torque, the state precesses away from the field direction and settles around the inverted state (yellow dot). Coloured dots on the sphere have the corresponding potential energy values indicated in \textbf{d}.}
  \label{fig:dynamicalstability}
\end{figure}

A rigid pendulum whose pivot is forced to vibrate up and down, known by the name of Kapitza~\cite{Kapitza1951}, has inspired generations of physicists~\cite{Acheson_1993,Citro_2015,Apffel_2020}. Under fast actuation, the bob appears to defy gravity by staying close to the upright position (Fig.~\ref{fig:dynamicalstability}a), demonstrating so-called dynamical stability~\cite{Stephenson_1908}. It represents a real-world example of a controllable non-linear dynamical system that exhibits a range of steady states in the long term, called attractors~\cite{Kuznetsov_1998}, such as complex periodic orbits~\cite{Acheson_1995} as well as chaotic motion~\cite{Blackburn_1998}. The fixed distance between the bob and the pivot drastically simplifies the equation of motion but preserves the non-linearity that causes the counter-intuitive dynamics.

Magnetic order shares the rigidity of a pendulum because the high exchange energy cost of modulating the saturation magnetisation $M_{\rm s}$ constrains the vector $\bm{M}$ to the Bloch sphere with radius $M_{\rm s}$. The Landau-Lifshitz-Gilbert (LLG) torque equation, a cousin of Newton's second law, relates the rate of change of the spin angular momentum density $-\gamma ^{-1}d\bm{M}/dt$ to the sum of various torques per volume $\bm{M}\times \delta _{\bm{M}}U +\bm{\tau }$  that act on the magnetisation. Here $\gamma >0$ is the modulus of the gyromagnetic ratio, $U$ is the magnetic free energy, $\delta _{\bm{M}}$ denotes the functional derivative with respect to $\bm{M}$, and $\bm{\tau }$ includes the viscous Gilbert damping~\cite{Gilbert_IEEE2004} and other torques that do not conserve $U$. Most studies focus on switching the magnetic order between the north and south poles of the Bloch sphere, which are the energy minima in the presence of a uniaxial anisotropy. This bistability is employed in magnetic memory applications~\cite{BHATTI_MaterToday2017} in which the two poles of a small magnetic medium represent a single bit that can be read out electrically by the celebrated giant/tunnelling magnetoresistance (MR)~\cite{Binasch_PRB1989,Baibich_PRL1988,Yuasa_NMater2004,Parkin_NMater2004}. The barrier between the energy minima is sufficiently higher than the thermal energy to retain the information safely~\cite{Ikeda_NMater2010,Yang_PRB2011,Yakata_JAP2009}. 

A static magnetic field tilts the free-energy double-well and favours one of the two minima into which the magnetisation settles in equilibrium (Fig.~\ref{fig:dynamicalstability}b). The magnetisation can be flipped by reversing the magnetic field, but also electrically using current-induced spin transfer from spin-polarised conduction electrons~\cite{SLONCZEWSKI_JMMM1996,Berger_PRB1996} that contributes to $\bm{\tau }$ without affecting $U$~\cite{Tserkovnyak_2005}. This effect, so-called spin transfer torque (STT)~\cite{RALPH_JMMM2008}, may counteract the magnetic dissipation that stabilises the magnetisation against fluctuations around a free energy minimum~\cite{Demidov_PRL2011}. When it overcomes the damping, the equilibrium becomes unstable, sending the system into motion, much like the oscillating pivot does to a rigid pendulum. In the presence of a large magnetic anisotropy, the switching is between the potential minima and non-volatile~\cite{Myers_Science1999,Mangin_NMater2006,Liu_Science2012} as illustrated in Fig.~\ref{fig:dynamicalstability}c. 
\note{Two decades ago, a volatile current-driven switching behaviour was observed once the applied magnetic field exceeds the anisotropy field~\cite{Katine_PRL2000}. It was subsequently identified to be a dynamical stabilisation at a potential maximum~\cite{Sun_PRB2000,Grollier_PRB2003,Ozyilmaz_PRL2003}(Fig.~\ref{fig:dynamicalstability}d), reminiscent of the Kapitza pendulum.}
\note{The physical mechanism is different, however, in the dissipative nature of STT contrasted to the effective conservative potential description of the driven pendulum where the inverted state is a local minimum~\cite{Kapitza1951}.}
\note{The high field and current density required for the dynamical stability in the hard magnetic layers in a nanopillar structure hindered further investigation of this regime. }
 
\rev{In this study, we employ magnetic films with vanishingly small magnetic anisotropy as a material platform to investigate dynamical stability and associated non-linear phenomena. As a model system, we use prototypical MgO|CoFeB|W multilayers in which the demagnetising field in the thin-film magnets favours CoFeB magnetisation within the film plane but can be counter-balanced by an interface-induced perpendicular magnetic anisotropy\cite{Ikeda_NMater2010,Yang_PRB2011,Yakata_JAP2009} (PMA) controlled by post-growth annealing (see Supplementary Note 1 for more details).}
We drive its non-linear dynamics by the STT generated by the spin-orbit interaction in the tungsten layer (spin-orbit torque, SOT)~\cite{Manchon_RevModPhys2019,Sinova_RevModPhys2015,Shao_TMAG2021}.
Two independent electrical measurements confirm that the STT forces the magnetisation into a steady state residing at the free energy maximum, contradicting the common wisdom that the attractor state under a large STT is an auto-oscillation, a particular kind of non-linear periodic orbit that appears as uniform and localised spin-wave modes~\cite{Tsoi_PRL1998,Kiselev_Nature2003,Demidov_NMater2012,Chen_IEEE2016,Mazraati_PRAppli2018,Ahlberg_BookChap2024}. A vanishing anisotropy suppresses auto-oscillations and drastically reduces the critical current to drive the magnet into the non-linear dynamical regime.
Our analytical estimates and numerical simulations (e.g.~Fig.~\ref{fig:dynamicalstability}f) show that the probability distribution of the magnetisation direction on the sphere can be controlled at will by the current and field. This whole Bloch sphere sampling might resource a continuous variable for probabilistic computation hardware, different from currently studied binary magnetic systems~\cite{Borders_Nature2019,Makiuchi_APL2021}.
Our devices serve as a foundation for studying non-linear dynamics of a complex many-body system that can be understood by a simple model.

\section*{Minimising the magnetic anisotropy}

We grew multilayer stacks of MgO(3~nm)|CoFeB(2~nm)|W(3~nm)
in an ultrahigh vacuum sputtering system on thermally oxidised Si at room temperature (see Methods). As-grown films are easy-plane magnets due to strong demagnetising fields of magnitude $M_\text{s}$. The PMA from the CoFeB|MgO interface~\cite{Meng_JAP2011,Aleksandrov_AIPAd2016} tends to pull the magnetisation out of plane which can be incorporated by an effective magnetisation $M_\text{eff} = M_\text{s} - 2K_{\perp}/(\mu_0 M_\text{s})$, where \(K_{\perp}\) is the perpendicular interface anisotropy constant and $\mu_0$ is the magnetic permeability of the vacuum. By careful growth and post-annealing we tune the system to $M_\text{eff}\approx 0 $ as explained in Supplementary Note 1. 

\begin{figure}[b!]
\centering
\includegraphics[width=1.02\linewidth]{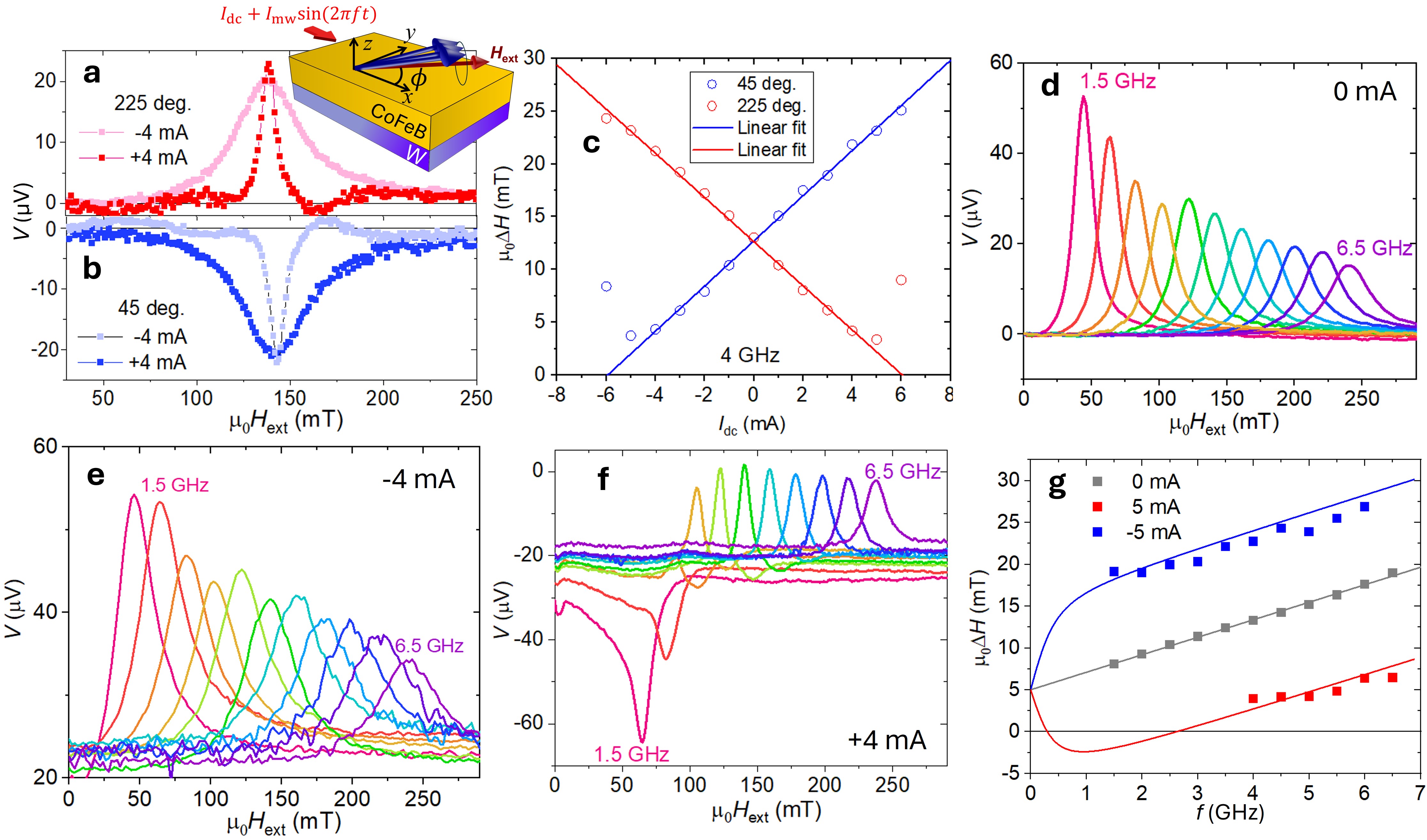}
 \caption{\textbf{Current-induced magnetic damping probed by ferromagnetic resonance.} \textbf{a-b}, Field-swept FMR voltages measured for (\textbf{a}) $\phi=225^{\circ}$ and (\textbf{b}) $\phi=45^{\circ}$. The inset is a sketch of our device and the coordinate system.  \textbf{c}, Linewidth extracted from FMR at 4 GHz with $\phi=45^{\circ}$ (blue) and  $\phi=225^{\circ}$(red), for different $I_\text{dc}$. Solid lines show the results of a linear fitting excluding points below -4 mA for 45$^{\circ}$ and above 4 mA for 225$^{\circ}$. \textbf{d-f}, Field-swept FMR voltages measured for various frequencies and three values of $I_\text{dc}$  (\textbf{d} 0 mA, \textbf{e} -4 mA and \textbf{f} 4 mA). Magnetic fields were applied along $\phi=225^{\circ}$. \textbf{g}, Linewidth as a function of frequency for different $I_\text{dc}$, i.e. +5, 0 and -5 mA, represented by red, grey and blue dots, respectively. The measurements were carried out at $\phi=225^{\circ}$. The grey line is a linear fitting of the 0 mA data. Both red and blue curves are calculated using Eq.~\ref{eq:linewidth} (see the main text).}
   \label{fig:stfmr}
\end{figure}

In the coordinate system in the inset of Fig.~\ref{fig:stfmr}a, the spin-Hall effect initiates the spin transfer by converting a charge current in the $x$ direction into a spin current normal to the interface and polarised along the $y$ direction~\cite{Liu_PRL2011,Manchon_RevModPhys2019}. The SOT arising from a dc current $I_{\rm dc}$ destabilises the equilibrium $\bm{M} \parallel \bm{H}_{\rm ext}$ at a critical value $I_{\rm c}$~\cite{Lee_APL2013,Sun_PRB2000,Katine_PRL2000} that solves
\begin{equation}
    -\beta I_{\rm c} \sin \phi = \alpha \left( H_{\rm ext}+\frac{M_{\rm eff}}{2} \right)  + \Delta H_0^{\prime }, \label{eq:critical}
\end{equation}
where $\alpha ,H_{\rm ext},\phi $ are the Gilbert damping constant, the external field magnitude, and the angle with the $x$ axis of the externally applied in-plane magnetic field, respectively. $\beta$ is a phenomenological parameter proportional to the spin Hall angle $\theta _{\rm SH}$ (see Methods) that characterises the efficiency of the charge-to-spin conversion. In Supplementary Note \rev{3} we explain that the current-induced self-torque in the CoFeB layer is negligibly small because electrons flow primarily in the W layer.  Spatial variations of the material parameters require the introduction of a constant offset $\Delta H_0^{\prime }$. We can reduce the critical current by minimising $M_{\rm eff}$, as expected.

\section*{Current rectification in the inverted state}

The dynamical stability of the anti-parallel state $\bm{M}=(M_x,M_y,M_z)^T\parallel -\bm{H}_{\rm ext}$ at large negative currents $I_{\rm dc}$ can be detected by measuring the spin-Hall magnetoresistance (SMR)~\cite{Nakayama_PRL2013,Chen_PRB2013} in static and dynamical regimes. 
The angular dependence of the electric resistance reads $R=R_0 + \Delta R_{\rm SMR} \left( 1- M_y^2 /M_\text{s}^2 \right) $ [\citeonline{Nakayama_PRL2013}] -- see detailed experiments of SMR in Supplementary Note \rev{3}. Here $R_0$ is the bulk resistance, while $\Delta R_{\rm SMR}$ can be calculated by spin-diffusion theory with appropriate boundary conditions~\cite{Chen_PRB2013} and/or fitted to the experiments. When a microwave current $I_{\rm mw}\cos\left( 2\pi ft \right) $ at frequency $f$ modulates $I_{\rm dc}$, the induced ac SOT causes a small oscillation \rev{$M_s \delta m_y $ }in $M_y$ that in turn affects the electric resistance. The resulting rectified voltage $V$ as a function of $f$ and $H_{\rm ext}$ is monitored by the spin-torque ferromagnetic resonance (ST-FMR)~\cite{Liu_PRL2011,Tulapurkar_Nature2005,Fang_NNano2011}. \rev{According to Supplementary Note 4, the resonance field and current-dependent line width are given by $H_{\rm res}^{\pm } = \sqrt{4\pi ^2 f^2 / (\gamma  \mu _0 )^2 + M_{\rm eff}^2 /4} \mp M_{\rm eff}/2$ and}
\rev{\begin{equation}
    \Delta H_{\pm } =  \Delta H_0 \pm \frac{2\pi f}{\gamma \mu _0} \left( \alpha + \frac{\beta I_{\rm dc}\sin \phi}{H_{\rm res}^{\pm }\pm M_{\rm eff}/2 } \right) ,\label{eq:linewidth}
\end{equation}}\rev{respectively, where the upper (lower) sign holds for the parallel (anti-parallel) state, and an inhomogeneous broadening $\Delta H_0$ is assumed independent of the magnetisation direction. To the leading order, $V \approx - 2\Delta R_{\rm SMR}I_{\rm mw} \left( M_y /M_s \right) \overline{\delta m_y \cos \left( 2\pi ft \right)}$ where the overbar denotes time averaging over a period. Since $\delta m_y \approx \beta I_{\rm mw}\cos \left( 2\pi ft\right) /\Delta H_{\pm }$ at $H_{\rm ext}=H^{\pm }_{\rm res}$ and $\Delta H_{\pm }>0 $ whenever $\bm{M}\parallel \pm \bm{H}_{\rm ext}$ is stable, $V$ changes sign on magnetisation reversal. With $M_y =\pm M_s \sin \phi $, the $H_{\rm ext}$ dependence of $V$ is given by}
\begin{equation}
    V =  \mp \Delta R_{\rm SMR}\beta  I_{\rm mw}^2 \frac{2\pi f/(\gamma \mu _0 )}{2H_{\rm res}^{\pm } \pm M_{\rm eff}}\frac{\Delta H_{\pm}\cos ^2 \phi \sin \phi }{\left( H_{\rm ext}-H_{\rm res}^{\pm } \right) ^2 + \Delta H_{\pm }^2}. \label{eq:voltage}
\end{equation} 
The Lorentzian approximation Eq.~(\ref{eq:voltage}) breaks down when $H_{\rm ext} \sim H_{\rm res}^{\pm }$ and $\Delta H_{\pm } \sim 0$ simultaneously (Supplementary Note \rev{4}).

\section*{Dynamical stability}

Figure~\ref{fig:stfmr} summarises our ST-FMR experiments on devices fabricated from magnetic multilayers with vanishing anisotropies as shown in Fig.~S1f. The FMR traces for $\phi=45^{\circ}$ and $225^{\circ}$ under applied currents $I_{\rm dc}=\pm 4$~mA at $f=4$ GHz in Figs.~\ref{fig:stfmr}a-b display clear linewidth changes~\cite{Ando＿PRL2008} as predicted by Eq.~(\ref{eq:linewidth}). The resonance fields $H_\text{res}$ and linewidths $\Delta H$ extracted by a double Lorentzian fit (see Methods) are plotted in Fig.~\ref{fig:stfmr}c and confirm the linear relationship between $\Delta H$ and $I_\text{dc}$ up to currents close to $I_\text{c}$ defined by Eq.~(\ref{eq:critical}). At higher currents, the system becomes increasingly unstable and $\Delta H$ increases again.  Linear extrapolation to zero $\Delta H$ gives $I_\text{c}\approx$~6~mA at $\mu_0 H_\text{ext}=\mu_0 H_\text{res}\approx 140$ mT or a current density of $2\times10^{11}$~Am$^{-2}$, which is approximately one order of magnitude smaller than the typical values for the onset of auto-oscillation in anisotropic magnets~\cite{Demidov_NMater2012,Duan_NComm2014}, due to the minimised $M_{\rm eff}$.
The deviation from the linear relation near $I_\text{c}$ reflects the aforementioned breakdown of Eq.~(\ref{eq:voltage}).

The FMR lineshape for $\phi=225^{\circ}$ and $I_\text{dc}$ of 0 and -4~mA in Figs.~\ref{fig:stfmr}d-e affirms the expected modulation of $\Delta H$, while fitting the field-frequency relation of the peak frequencies by the Kittel formula yields the residual easy-plane anisotropy of $\mu_0  M_{\rm eff} =37.3$ mT, which is indeed much smaller than $\mu_0 M_\text{s}$ of CoFeB ($\thicksim$1.5~T)~\cite{Kim_SciRep2022}. At $I_{\rm dc}=4$~mA in Fig.~\ref{fig:stfmr}f, $V$ changes its sign at low frequencies, which is strong evidence of the dynamical stabilisation of the anti-parallel state $\bm{M}\parallel -\bm{H}_{\rm ext}$ for $\mu_0 H_{\rm ext} \lesssim 100$ mT. This also agrees with Eq.~(\ref{eq:critical}) where $I_{\rm c}=4$ mA  for $\phi = 225^{\circ}$ and $\mu_0 H_{\rm ext}=100$ mT.  The fit of the linear relation in Fig.~\ref{fig:stfmr}c leads to $\theta_\text{SH}=0.22$ which is similar to published values~\cite{Pai_APL2012} and $\beta=2.6\times10^6$ m$^{-1}$.
We confirm the sign reversal of the ST-FMR peaks for small $f$ and $H_{\rm ext}$ in different samples as shown in Supplementary Fig.~\rev{S10}. Equation~(\ref{eq:linewidth}) offers a quantitative test of our interpretation. For $2\pi f/\gamma  \gg \left| M_{\rm eff} \right|$ and $H_{\rm res}^+ +M_{\rm eff}/2 \propto f$, $\Delta H$ becomes a linear function of $f$ with a slope $2\pi \alpha /\gamma \mu _0 $ that does not depend on $I_{\rm dc}$. Figure~\ref{fig:stfmr}g verifies this prediction, since $I_{\rm dc}$ mainly produces the $\Delta H$ offset \rev{of} $2\pi f \beta I_{\rm dc}\sin \phi /\sqrt{4\pi^2 f^2 +\gamma ^2 \mu _0^2 M_{\rm eff}^2 /4}$ for $2\pi f\gg \gamma \mu _0 M_{\rm eff}$, in agreement with the curves that represent Eq.~(\ref{eq:linewidth}) for the upper signs with $\alpha =0.053$ and $\mu_0 \Delta H_0 = 5$ mT.

\section*{Electric resistance probes magnetic fluctuations}

\begin{figure}[b!]
\centering
\includegraphics[width=1\linewidth]{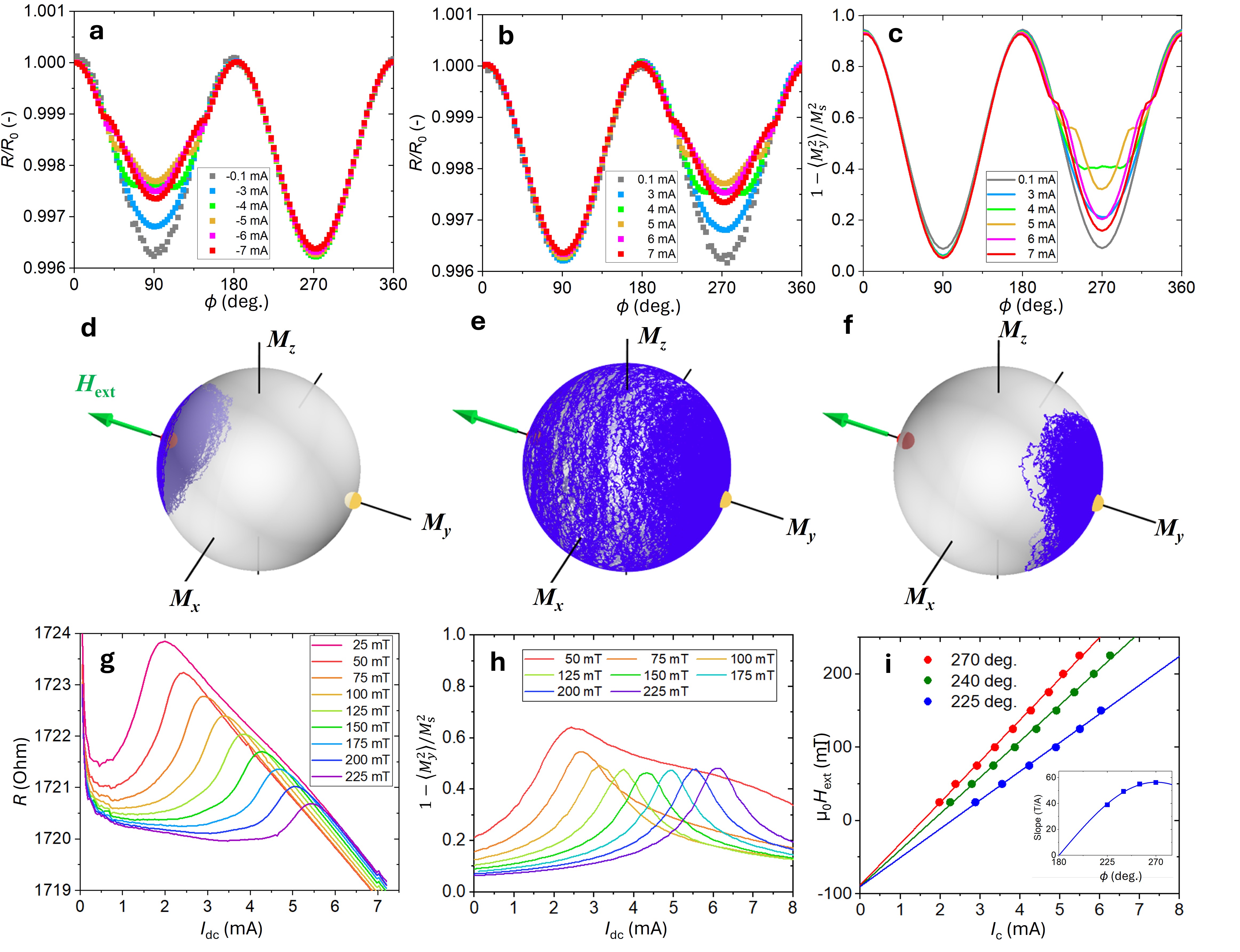}
\caption{\textbf{Electric resistance probes for magnetic fluctuations and the inverted states.} \textbf{a-b}, Observed angular MR in the coordinate system of Fig.~\ref{fig:stfmr}a for negative (\textbf{a}) and positive (\textbf{b}) currents, normalised by the resistance at $\phi =0^{\circ }$ for each current. \textbf{c}, $1- \langle M_y^2 \rangle /M_\text{s}^2 $ calculated by the stochastic LLG equation in the macrospin approximation. \textbf{d-f}, Time-dependent trajectories of the macrospin moment calculated by the stochastic LLG equation for currents 0.1~mA (\textbf{d}), 4.1~mA (\textbf{e}) and 7~mA (\textbf{f}), and $\phi=270^{\circ}$. The field strength is 125~mT for panels \textbf{a}-\textbf{f}.  \textbf{g}, Resistance vs. dc current measurements for applied magnetic fields 25-200~mT along $\phi=270^{\circ}$. \textbf{h},  Calculated $1- \langle M_y^2\rangle /M_{\rm s}^2$ as a function of current and applied magnetic field. \textbf{i}, $I_\text{c}$ extracted in \textbf{g} for different field strengths and angles (dots) and linear fits (lines). The inset shows the slope (dots) and a fit by $\sin\phi$ (curve).}
  \label{fig:mr1}
\end{figure}

The thermal fluctuations of the magnetisation angle affect its time-averaged properties. In particular, when the stability of both poles near the critical current weakens, the magnetisation becomes susceptible to even small random torques. Because the time scales associated with the motion of conduction electrons (fs) are much faster than those of magnetic fluctuations (ns), the SMR probes the expectation value of $M_{y}^2$, i.e. $R= R_0 + \Delta R_{\rm SMR}\left( 1- \langle M_y^2 \rangle /M_{\rm s}^2 \right) $ in which the angled brackets denote temporal and spatial averaging~\cite{Nakayama_PRL2013}. Larger fluctuations of $\bm{M}$ imply smaller $\langle M_y^2 \rangle $ and a higher resistance in the parallel state \rev{(see more discussions in relation to unidirectional SMR\cite{Avci_PRL2018,Borisenko_APL2018} in Supplementary Note 3).}

Figures~\ref{fig:mr1}a-b show the $\phi$ dependence of the MR for $\mu_0 H_{\rm ext} = 150$~mT and different $I_\text{dc}$. Without fluctuations, $R/R_0 =1+ \left( \Delta R_{\rm SMR}/R_0 \right) \left( 1-\sin ^2 \phi \right) $, which agrees well for small applied currents $I_{\rm dc}=\pm 0.1$ mA. Joule heating cannot explain the asymmetric distortion of the MR around $\phi =180^{\circ}$ at larger $\left| I_\text{dc} \right| $. A strong current dependence occurs only when the field direction and spin polarisation are parallel, i.e., when the STT destabilises the equilibrium magnetisation. Beyond currents of 4 and 5 mA, the fluctuations appear to decrease again.

We model the magnetisation fluctuations by a stochastic LLG equation that includes the thermal noise in the effective magnetic field $\bm{\xi }$ that obeys the fluctuation-dissipation relation $\langle \xi _i \left( t\right) \xi _j \left( t^{\prime } \right) \rangle = \delta _{ij} \delta \left( t-t^{\prime } \right) \left( 2\alpha k_\text{B} T\right) /\left( \gamma (1+\alpha^2) M_{\rm s} V_\text{a}\right)$, where $k_\text{B}$ is the Boltzmann constant and $V_\text{a}=1\times 25\times 25$~nm$^3$ is a magnetic coherence volume~\cite{Xiao_PRB2010}.
The calculated $1-\langle M_{y}^2 \rangle /M_{\rm s}^2 $ as a function of $\phi $ for fixed $I_{\rm dc}$ in Fig.~\ref{fig:mr1}c reproduces the salient trends observed in Fig.~\ref{fig:mr1}b.  $\bm{M}(t)$ for current values $I_{\rm dc}=$ 0.1, 4.1 and 7~mA and $\phi = 270^{\circ}$ in Figs.~\ref{fig:mr1}d-f illustrate the build-up of the steady state probability distributions in Fig.~\ref{fig:mr1}c. For very small and large currents, at least at one of the poles the torques damp rather than amplify the thermal agitation. Close to the critical current (Fig.~\ref{fig:mr1}e), the magnetisation vector explores the entire Bloch sphere with nearly constant probability density.

\begin{figure}[b!]
\centering
\includegraphics[width=1\linewidth]{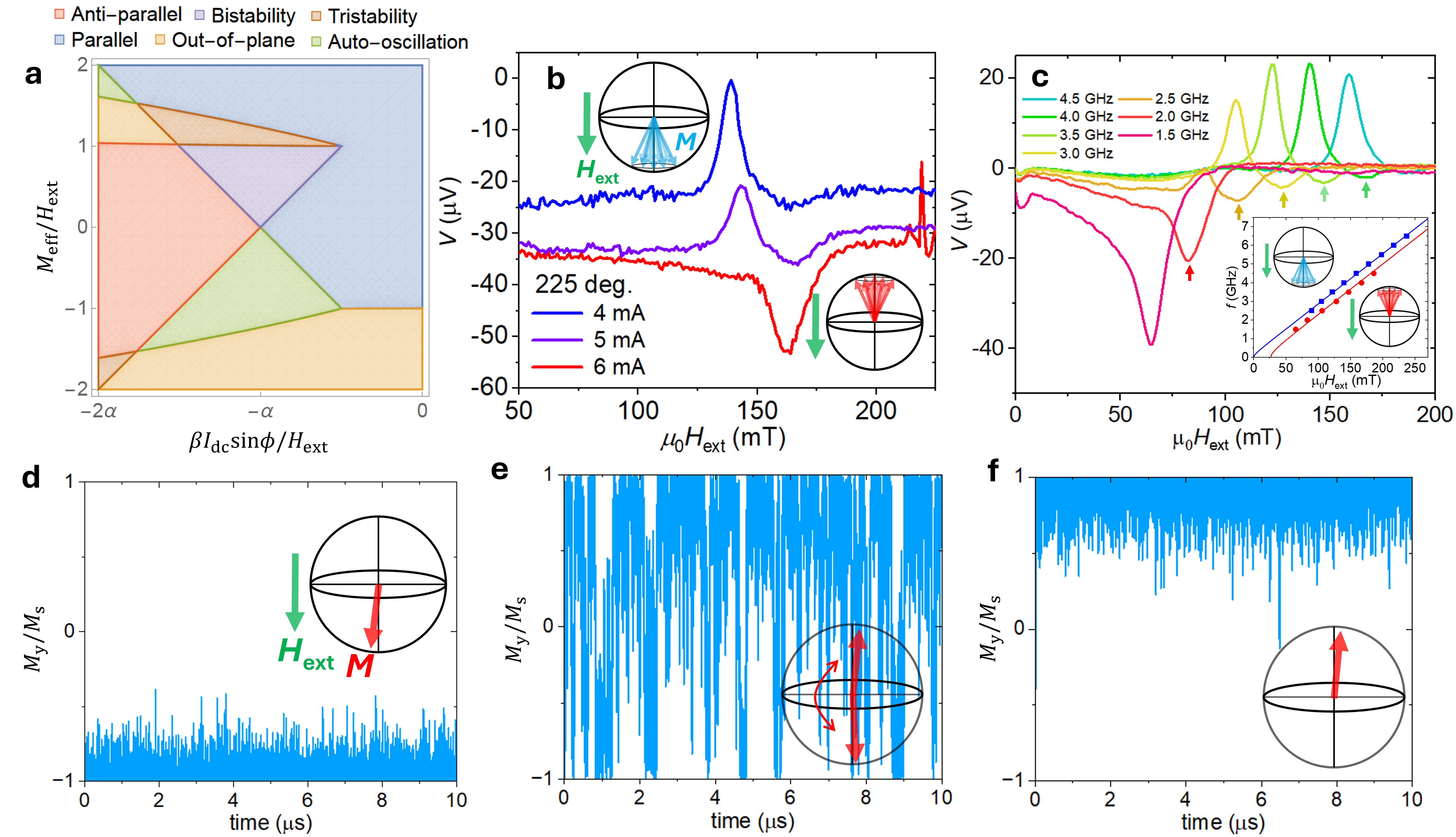}
\caption{\textbf{Stability diagram, bistability and probabilistic distribution for nearly isotropic magnets.} \textbf{a}, Bifurcation diagram constructed from the stationary solutions of the LLG equation. \textbf{b}, ST-FMR for different $I_\text{dc}$ close to the damping compensation condition. When a larger $I_\text{dc}$ increases the anti-damping torques beyond the critical value, the magnetisation precesses around the dynamically stable state at the north pole, generating an opposite dc voltage at resonance. \textbf{c}, The ST-FMR results highlight the current-induced magnetisation reversal. The inset shows the FMR frequencies as functions of field for the two magnetisation directions. \textbf{d-f}, Time-dependent traces of the normalised $M_y$ component for small(\textbf{d})/medium(\textbf{e})/large(\textbf{f}) anti-damping torques. Here $M_y=-1$ is along the magnetic field direction. }
  \label{fig:phase}
\end{figure}

Equation~(\ref{eq:critical}) should hold near the resistance maxima in Fig.~\ref{fig:mr1}g that shift towards higher $I_\text{dc}$ for larger $H_\text{ext}$, but disappear when switching the field direction as shown in Supplementary Fig.~\rev{S12a}.  At $\mu_0 H_{\rm ext}=150$ mT the resistance peak height relative to lower $I_\text{dc}$ is about 1-2~$\Omega$ or 0.1~\%, consistent with Fig.~\ref{fig:mr1}b at $\phi$ = 270$^{\circ}$. The peaks in the calculated $1- \langle M_y^2 \rangle /M_{\rm s}^2$ in Fig.~\ref{fig:mr1}h correspond to a minimum in $\langle M_{y}^2 \rangle /M_{\rm s}^2$ or maximum of the magnetic fluctuations at $I_{\rm c}$. \rev{The differences between theory and experiments (Figs.~\ref{fig:mr1}g and ~\ref{fig:mr1}h) are partially caused by the incomplete theoretical account of heating effects, as well as the limitation of the macrospin model to describe spatial fluctuations in our extended film geometry.} Equation~(\ref{eq:critical}) predicts a linear relationship between $H_{\rm ext}$ and $I_{\rm c}$ with a negative intercept. Figure~\ref{fig:mr1}i illustrates the linear field-current relationship observed at the resistance maximum. A fit to the $\phi$ dependence of the slope (inset of Fig.~\ref{fig:mr1}i and Fig.~\rev{S11e}) leads to the ratio $\theta_\text{SH}/\alpha = 3.7$. Substituting $\alpha=0.053$ from the ST-FMR, $\theta_\text{SH}=0.20$  agrees well with $\theta_\text{SH}=0.22$ found above from an independent experiment on the same sample. Using Eq.~(\ref{eq:critical}) with the experimental parameters $\alpha$ and $M_{\rm eff}$ and the zero-current intercept in Fig.~\ref{fig:mr1}i, $\mu_0 \Delta H_0^{\prime} = 3.8$~mT, comparable to the inhomogeneous broadening $\mu_0 \Delta H_0 = 5.0$~mT as deduced from the ST-FMR.

\section*{Stochastic switching in nearly isotropic magnets}

Both ST-FMR and MR measurements find a residual easy-plane anisotropy $M_{\rm eff}>0$ that, albeit small, causes an effect on the non-linear dynamics that deserves a more detailed study. Figure~\ref{fig:phase}a shows a bifurcation (or dynamical phase) diagram that analytically classifies the asymptotic final states of the deterministic LLG equation as a function of dimensionless parameters $\beta I_{\rm dc}\sin \phi /H_{\rm ext}$ and $M_{\rm eff}/H_{\rm ext}$ (see more details in Supplementary Note \rev{5}). Since the Poincar{\'e}-Bendixon theorem~\cite{Teschl_book2012} excludes chaos on a sphere, the attractors in the absence of stable time-independent solutions must be finite-angle precessional states. The ``auto-oscillations"  that occupy the regions shaded in green exist for $M_{\rm eff}>0$ only when the anisotropy and STT are very large. We identify a regime around the damping compensation $-\beta I_{\rm dc}\sin \phi /H_{\rm ext}=\alpha $, in which the inverted state is dynamically stabilised by a current that is smaller than the estimate Eq.~(\ref{eq:critical}) for  $M_{\rm eff}=0$. We may explain this deviation in terms of a dynamical bistability, also observed in the simulations in Fig.~\ref{fig:mr1}e in the form of a bimodal probability density at the critical current.

The experiments show clear signatures of this bistability. The ST-FMR voltages in Fig.~\ref{fig:phase}b change sign with increasing $I_\text{dc}$. The FMR mode around $\bm{M} \parallel \bm{H}_{\rm ext}$ at $\mu _0 H_\text{ext}\approx 140$~mT gradually disappears, and an inverted peak emerges at a higher field $\mu _0 H_{\rm ext} \approx 160$~mT. This observation supports the prediction of different resonance fields $H^+_{\rm res}-H^-_{\rm res} =-M_{\rm eff}$ in Eq.~(\ref{eq:voltage}). At $I_{\rm dc}=5$~mA, two peaks are observed. The one at $\mu _0 H_{\rm ext}\approx 140$~mT for $\bm{M}\parallel \bm{H}_{\rm ext}$ is according to Eq.~(\ref{eq:critical}) stable at higher fields. The inverted peak at $\mu _0 H_{\rm ext}\approx 160 >140$~mT, indicates the need for two independent stability criteria. Figure~\ref{fig:phase}c also shows two peaks in ST-FMR voltages at different frequencies and a fixed current of $I_{\rm dc}=4$ mA for $100~\textrm{mT}\lesssim \mu _0 H_{\rm ext}\lesssim 170~\textrm{mT}$. Theory predicts different resonance conditions for the parallel and inverted states $2\pi f_{\rm res}^{\pm }=\gamma \mu _0 \sqrt{H_{\rm ext}\left( H_{\rm ext}\mp M_{\rm eff}\right)}$ for $\bm{M}\parallel \pm \bm{H}_{\rm ext}$ respectively, consistent with $H_{\rm res}$ in the inset of Fig.~\ref{fig:phase}c, which allows us to estimate $\mu _0 M_{\rm eff} \approx 30 \pm 5$~mT. 

\begin{figure}[b!]
\centering
\includegraphics[width=\linewidth]{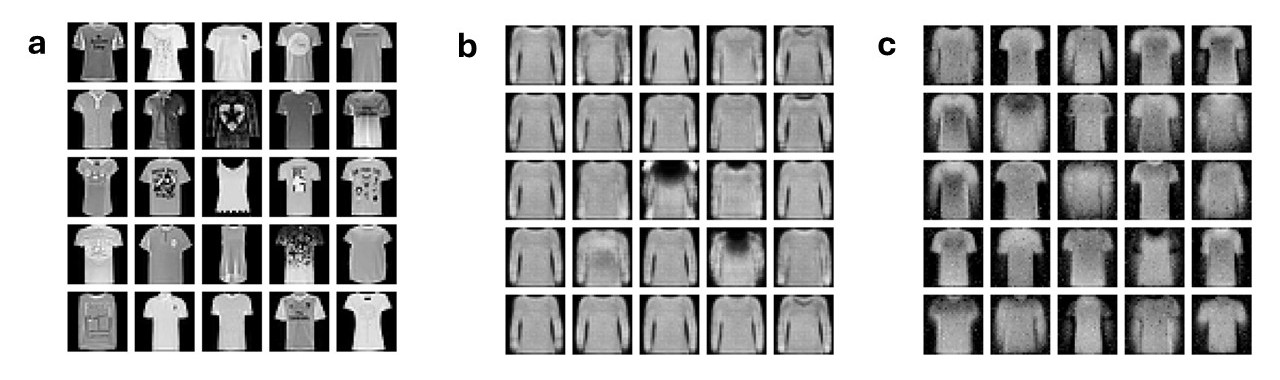}
\caption{\rev{\textbf{Two-dimensional image generation by a continuous restricted Boltzmann machine with electrically controlled zero-damping states in isotropic magnets. a,} Samples of the Fashion-MNIST training data (t-shirt class). \textbf{b \& c,} Generated data using binary and continuous variables for the visible nodes respectively.}}
\label{fig:RBM}
\end{figure}

Our discovery of dynamically stabilised states opens a vast and unexplored field of research. The numerical solutions of the stochastic LLG equation for $M_{\rm eff}/H_{\rm ext} = 0.31$ in Figs.~\ref{fig:phase}d-f show drastic changes of the probabilistic distribution when relaxing the constraint of a strong anisotropy. At an intermediate current $I_{\rm dc}=4.0$~mA, the magnetisation visits all angles of the Bloch sphere with comparable probability. 
The characteristic dwell time around the poles in Fig.~\ref{fig:phase}e is in the sub-microsecond range, consistent with a semi-analytical estimate from the ``first-passage time" theory~\cite{Newhall2013,Taniguchi2013} (Supplementary Note \rev{7}). It is predicted to be highly tunable through an exponential dependence on a Boltzmann factor $\mu _0 M_{\rm s} H_{\rm ext}V_\text{a}/k_B T$, which can be probed by time-resolved optical or electric measurements. An analytical solution of the Fokker-Planck equation corresponding to the stochastic LLG for $M_{\rm eff}=0$ gives the same exponential dependence on $\mu _0 M_{\rm s} H_{\rm ext}V_\text{a}/k_B T$ of the MR vs. $I_{\rm dc}$ (Supplementary Note \rev{6}). Figure \rev{S12}, on the other hand, shows that the current dependence of the MR is remarkably robust across a wide range of temperature. \rev{We attribute the discrepancies to spatial fluctuations in the film beyond the macrospin model, such as sample inhomogeneities, domain formation, or spin-wave instabilities of the anti-parallel state~\cite{Harms_AIPAdv2024}.} Furthermore, our understanding of the crystalline, compositional, and thickness inhomogeneities and their effects on the critical currents is incomplete, as is our interpretation of $\Delta H_0^{\prime }$ in Eq.~(\ref{eq:critical}).  

\rev{Electrically controlled zero-damping states in isotropic magnets provide a computational resource that samples continuous random variables from the entire Bloch sphere, physically modelling a continuous restricted Boltzmann machine~\cite{salakhutdinov2007rbm,bereux2025rbm} (RBM). We use this model to generate two-dimensional images via an unsupervised machine learning framework. In this scheme, we define visible and hidden layers connected by trainable weights, and apply individual biases on each node (see Supplementary Note 9 for more details). The continuous nodes in the visible layer are represented by the magnetisation dynamics of an STT-driven isotropic magnet. We trained the continuous RBM and a traditional binary RBM using the Fashion-MNIST benchmark images (Fig.~\ref{fig:RBM}a) and at inference, generated new samples, as shown in Figs.~\ref{fig:RBM}b and~\ref{fig:RBM}c. 
The generative artificial intelligence architecture using the continuous variable outperforms those using binary variables
by producing more meaningful and diverse images, which we quantitatively demonstrate using standard metrics as detailed in Supplementary Note~9. This improvement arises because, unlike the traditional RBM which is limited to generating binary image data and necessitates averaging during post processing (rate-coded RBM)~\cite{chen2002crbm,chen2003crbm}, the continuous RBM directly generates raw image data, enabling smoother convergence and more diverse image generation. Our findings highlight the potential of these magnetic states for various computing applications, superseding conventional binary systems such as stochastic magnetic tunnel junctions acting as probabilistic bits~\cite{borders2019smtjs}.}

To summarise, nearly isotropic MgO|CoFeB|W multilayers allow a dynamical stabilisation of the magnetic order at the maximum of the free energy simply by applying an electric current, creating a research frontier for non-linear and strongly out-of-equilibrium magnetic properties. \note{Building on our} time-averaged measurements on macroscopic samples \note{that} reveal the essential physics of this phenomenon, spatially and temporally resolved experiments can test phenomenological, statistical, and ultimately quantum-mechanical models. \note{In this regard, microfocus Brillouin light scattering spectroscopy would be a powerful tool\cite{Demidov_PRL2011}}. 
The dynamical stabilisation of magnetic order has not yet played a major role in studies of magnonic and spintronic effects\cite{Kruglyak_Magnonics2010,Chumak_NPhys2015,Barman_IOP2021}. However, the option of creating magnetic devices based on isotropic magnets that operate in inverted and dynamically bistable regimes at relatively low power levels provides radically different functionalities that may be useful for, e.g. probabilistic computation hardware~\cite{Chowdhury_IEEE2023}. A robust dynamically stabilised state \rev{in the extended film geometry} can also lead to experimental studies of recent theoretically predicted antiparticle-like (or anti-magnonic) properties of spin waves around the reversed magnetisation state~\cite{Harms_AIPAdv2024}. 

\section*{Methods}

\noindent \textbf{Definition of $\beta$} 

\noindent We parametrise the efficiency of current-induced STTs in the planar SOT geometry by the fit parameter $\beta=\hslash \theta_{\mathrm{SH}}/(2 e \mu_0 M_\text{s} w d_{\mathrm{M}} d_{\mathrm{W}})$, where $\hbar$, $e$, $w$, $d_\text{M}$ and $d_\text{W}$ are the reduced Planck constant, the elementary charge, the device bar width, the thicknesses of CoFeB and W layers, respectively.  \\

\noindent \textbf{Thin film preparation, post-annealing and device fabrication} 

\noindent Thin films were grown on thermally-oxidised Si (Si-O/Si) substrates by magnetron sputtering in an ultrahigh vacuum apparatus with the base pressure below 2 × 10$^{-7}$ Pa. 
The stacking of the thin films was Si substrate|Si-O|W (3 nm)|CoFeB (2 nm)|MgO (3 nm). 
All layers were deposited at ambient temperature under the Ar gas pressure of $\approx$ 5 mTorr. The W and CoFeB (atomic composition of Co$_{20}$Fe$_{60}$B$_{20}$) targets were sputtered using dc power supplies while the MgO target was sputtered using an rf power supply. The deposition rates were set to 0.029~nm~s$^{-1}$ for W, 0.022~nm~s$^{-1}$ for CoFeB and 0.0016~nm~s$^{-1}$ for MgO. \rev{The structural characterisation of one of the post-annealed W|CoFeB|MgO sample was carried out by scanning transmission
electron microscopy (STEM) and shown in Supplementary Note 2.}
After each sputter-deposition, the films were transferred to the vacuum furnace with the base pressure below 1 × 10$^{-4}$~Pa in which the films were annealed at a target temperature for one hour -- see the annealing condition in Supplementary Note 1. 
The thin films were patterned into a rectangle with 10~$\mu$m width and 40~$\mu$m length by Ar ion milling using standard optical lithography, onto which Au electrodes patterned into a coplanar waveguide were made by sputtering and lift-off techniques. \\

\noindent \textbf{ST-FMR and MR measurements} 

\noindent The devices were measured in a microwave prober system to minimise the losses between a microwave signal generator and the device. Typically, we inject microwaves with a power of 13~dBm by the signal generator, with amplitudes modulated by 25~\% at a frequency of 10~kHz. For ST-FMR measurements, this $I_{\rm mw}$sin(2$\pi ft$) induces a rectification voltage across the device that is measured through a bias-tee using a lock-in amplifier set to 10~kHz. $I_\text{dc}$ is applied through the bias-tee and a 10~k$\Omega$ resistor. The measured data of $V$ are fitted using the following functions.
\begin{equation*}
    V =  V_0 + V_\text{sym} \frac{\Delta H^2}{(H_\text{ext}-H_\text{res})^2+\Delta H^2} + V_\text{asy}\frac{\Delta H(H_\text{ext}-H_\text{res})}{(H_\text{ext}-H_\text{res})^2+\Delta H^2},\nonumber
\end{equation*}
where $V_0$, $V_\text{sym}$ and $V_\text{asy}$ are the field-independent voltage, the amplitude of the symmetric and antisymmetric components around $H_\text{res}$. We extract $H_\text{res}$ for Fig.~\ref{fig:phase}c by using $H_\text{ext}$ that maximised and minimised the peak amplitude. We measured MR by applying $I_\text{dc}$ through the 10~k$\Omega$ resistor in series.  
\\

\newpage

\noindent \textbf{Macrospin simulations} 

\noindent The Landau-Lifshitz-Gilbert equation with a damping-like spin torque term reads:
\begin{equation}
    \frac{\mathrm{d}\bm{n}}{\mathrm{d}t} = -\frac{\gamma}{1+\alpha^2}\bm{n}\times\bm{B}^{\mathrm{eff}} -\frac{\gamma\alpha}{1+\alpha^2}\bm{n}\times\left(\bm{n}\times\bm{B}^{\mathrm{eff}}\right) -\gamma\beta I_{\rm dc} \bm{n}\times\left(\bm{n}\times\bm{\sigma}\right),
\end{equation}
where $\bm{n}$ is the dimensionless unit vector of the magnetisation, $\bm{B}^{\mathrm{eff}}$ is the effective field in Tesla, $\alpha=0.09$
%\kei{Please confirm.}\joe{Correct}
is the dimensionless Gilbert damping, $\gamma=1.76\times10^{11}$~rad~$s^{-1}$~T$^{-1}$ is the gyromagnetic ratio and $\bm{\sigma}$ is the direction of the electron spin polarisation The spin transfer torque has a strength $\beta j$ with $\beta=\hslash \theta_{\mathrm{SH}}/(2 e \mu_0 M_\text{s} w d_{\mathrm{M}} d_{\mathrm{W}})=2.6\times 10^6 $ where $\theta_{\mathrm{SH}}=0.22$ is the spin Hall angle, $M_\text{s}=758$~kA~m$^{-1}$ is the saturation magnetisation of CoFeB, $w=10$~$\mu$m is the width of the sample (perpendicular to the current direction), $d_{\mathrm{W}}=3$~nm is the thickness of the normal metal and we assume all of the electronic current flows through this layer due to the higher conductivity. $d_{\mathrm{M}}=1$~nm is the thickness of the magnetic layer. We integrate the equation of motion numerically using a fourth-order Runge-Kutta scheme with a time-step $\Delta t=0.01$~ps. The effective field contains terms for the applied field, the demagnetising effect of a thin film and a PMA i.e.
\begin{equation}
    \bm{B}^{\mathrm{eff}} = \mu_0\bm{H}_{\text{ext}} - \left(\mu_0 M_{\rm s} - \frac{2K_{\perp}}{M_{\rm s}}\right)\vec{e}_{z} +\bm{\xi},
\end{equation}
where $\mu_0$ is the vacuum permeability, $\mu_0\bm{H}_{\text{ext}}$ is the externally applied magnetic field in Tesla, $K_{\perp}=327$~kJ~m$^{-3}$ is the magneto-crystalline anisotropy energy density and $\vec{e}_{z}$ is a unit vector along the $z$-direction. $\bm{\xi}$ is a stochastic thermal field in Tesla with white noise characteristics governed by the fluctuation dissipation theorem $\langle \bm{\xi}\rangle = 0; \langle \xi(t)\xi(t') \rangle = \delta(t-t')(2\alpha k_\text{B} T)/(\gamma (1+\alpha^2) M_\text{s} V_\text{a})$ where $V_\text{a}=1\times25\times25$~nm$^3$ is the effective volume of a macrospin region in the extended film. We include the effect of Joule heating by renormalising the anisotropy and magnetisation at a given temperature. More details are given in Supplementary Note \rev{8}.

\section*{Data availability}

The data presented in the main text and the Supplementary Information are available from the corresponding authors upon reasonable request.

\bibliography{references_switching}% common bib file

\section*{Acknowledgements}

The authors thank Philipp Pirro, Abbas Koujok, Mathias Kl{\"a}ui, Olivier Klein, Christian Back, Lin Chen, Eiji Saitoh, Kerem Çamsarı, and Tomohiro Taniguchi for stimulating discussions. H.K. thanks the Leverhulme Trust for financial support via their research fellowship (RF-2024-317), and the Engineering and Physical Sciences Research Council (EPSRC) for support via grant EP/X015661/1. J.B. acknowledges funding from a Royal Society University Research Fellowship. H.Y. thanks the EPSRC for funding through the EPSRC DTP studentship (EP/T517793/1 and EP/W524335/1). D.P. is supported by the EPSRC and SFI Centre for Doctoral Training in Advanced Characterisation of Materials Grant Ref: EP/S023259/1. This project was supported by JSPS KAKENHI (Nos.~21K13886, 24K00576, JP23H00232, 22H04965, JP24H02231, JP25K17931 \rev{and 25H00837}), X-NICS of MEXT (No. JPJ011438) and JST PRESTO (Grant No.~JPMJPR20LB). Part of this work was undertaken on the Aire HPC system at the University of Leeds, UK. The device fabrication was partly carried out at the Cooperative Research and Development Center for Advanced Materials, Institute for Materials Research, Tohoku University. The authors thank T. Sasaki for their Au film deposition. This work was partially supported by the Collaborative Research Center on Energy Materials, Institute for Materials Research (E-IMR), Tohoku University. H.K. and T.S. acknowledge support by the GIMRT Program of the Institute for Materials Research, and by the Invitational Fellowship Program for Collaborative Research with International Researcher 2024, Tohoku University. J.B. and K.Y. acknowledge support from a Royal Society International Exchange 2023 Cost Share (JSPS)/JSPS Bilateral Program Number JPJSBP120245708.

\section*{Author Contributions}
H.K., T.S. and K.Y. started this study in summer 2022, supported by the visiting professorship programme at Institute for Materials Research, Tohoku University. H.K., T.S. and V.K.K. grew thin-films and fabricated them into micro-devices. H.K., T.S., T.Y.\rev{, T.D., D.P.} and K.D.S. performed magnetic and transport measurements presented in this work. H.K., T.S.\rev{, T.D., D.P.} and K.Y. carried out data analysis, supported by all the co-authors. K.Y. theorised the dynamical stability analysis. K.Y. and G.E.W.B. provided critical insights for the early state of this study. J.B. performed the macro-spin simulations and contributed to finalise this study. H.Y. performed the continuous restricted Boltzmann machine work supervised by H.K. \rev{T.Y. and X.H. performed the structural analysis by STEM.}  H.K., K.Y., J.B., G.E.W.B. and T.S. prepared the manuscript where the rest provided comments.  

\newpage
\makeatother
\makeatletter
\renewcommand \thefigure{S\@arabic\c@figure}
\makeatother
\renewcommand{\theequation}{S\arabic{equation}}
\renewcommand{\vec}[1]{{\ensuremath{\bm{\mathrm{#1}}}}}

\vspace{5 mm}

\section*{{\huge Supplementary Information: \\Dynamical stability by spin transfer in nearly isotropic magnets}}

\vspace{3 mm}

\noindent {\Large Hidekazu Kurebayashi$^{1,2,3,4,5,*}$, Joseph Barker$^{6}$, Takumi Yamazaki$^4$, Varun K. Kushwaha$^4$, Kilian D. Stenning$^7$, Harry Youel$^1$, Xueyao Hou$^4$, Troy Dion$^{3,4}$, Daniel Prestwood$^1$, Gerrit E. W. Bauer$^{3,4,5,8}$, Kei Yamamoto$^{9,*}$, and Takeshi Seki$^{3,4,*}$}

\vspace{5 mm}

\noindent $^1${\large London Centre for Nanotechnology, University College London,  17-19 Gordon Street, London, WC1H 0AH, United Kingdom}
\vspace{1 mm}

\noindent $^2${\large Department of Electronic and Electrical Engineering, University College London, Roberts Building, London, WC1E 7JE, United Kingdom}
\vspace{1 mm}

\noindent $^3${\large Center for Science and Innovation in Spintronics, Tohoku University, 2-1-1, Katahira, Sendai, 980-8577 Japan}
\vspace{1 mm}

\noindent $^4${\large Institute for Materials Research, Tohoku University, 2-1-1, Katahira, Sendai, 980-8577 Japan}
\vspace{1 mm}

\noindent $^5${\large WPI Advanced Institute for Materials Research, Tohoku University, 2-1-1, Katahira, Sendai 980-8577, Japan}
\vspace{1 mm}

\noindent $^6${\large School of Physics and Astronomy, University of Leeds, Leeds LS2 9JT, United Kingdom}
\vspace{1 mm}

\noindent $^7${\large Blackett Laboratory, Imperial College London, London SW7 2AZ, United Kingdom}
\vspace{1 mm}

\noindent $^8${\large Kavli Institute for Theoretical Sciences, University of the Chinese Academy of Sciences, Beijing 100190, China}
\vspace{1 mm}

\noindent $^9${\large Advanced Science Research Center, Japan Atomic Energy Agency, 2-4 Shirakata, Tokai 319-1195, Japan}
\vspace{1 mm}

\noindent $^*${\large Corresponding author e-mail: h.kurebayashi@ucl.ac.uk, yamamoto.kei@jaea.go.jp, takeshi.seki@tohoku.ac.jp}

\vspace{12 mm}

\begin{center} {\Large\textbf{CONTENTS}} \end{center}
{\large Supplementary Note 1: Minimising the Magnetic Anisotropy}\vspace{0.2cm} \\
{\large \rev{Supplementary Note 2: Scanning transmission electron microscopy characterisation}}\vspace{0.2cm} \\
{\large Supplementary Note \rev{3}: magnetoresistance (MR)}\vspace{0.2cm} \\
{\large Supplementary Note \rev{4}: Theoretical Modelling of ST-FMR under Dynamical Stabilisation}\vspace{0.2cm}\\
{\large Supplementary Note \rev{5}: Stability Diagram}\vspace{0.2cm}\\
{\large Supplementary Note \rev{6}: Fokker-Planck Equation}\vspace{0.2cm}\\
{\large Supplementary Note \rev{7}: First Passage Time}\vspace{0.2cm}\\
{\large Supplementary Note \rev{8}: Joule Heating}\vspace{0.2cm}\\
{\large Supplementary Note \rev{9}: Continuous restricted Boltzmann machines using nearly isotropic magnets}\vspace{0.2cm}\\
{\large Further Supplementary Figures \rev{S10-S14}}

% reset the counter
\setcounter{figure}{0}  % next figure will be 1 again

\newpage
\setcounter{equation}{0}

\section*{Supplementary Note 1: Minimising the Magnetic Anisotropy}

\begin{figure}[bh]
    \centering
    \includegraphics[width=1.0\linewidth]{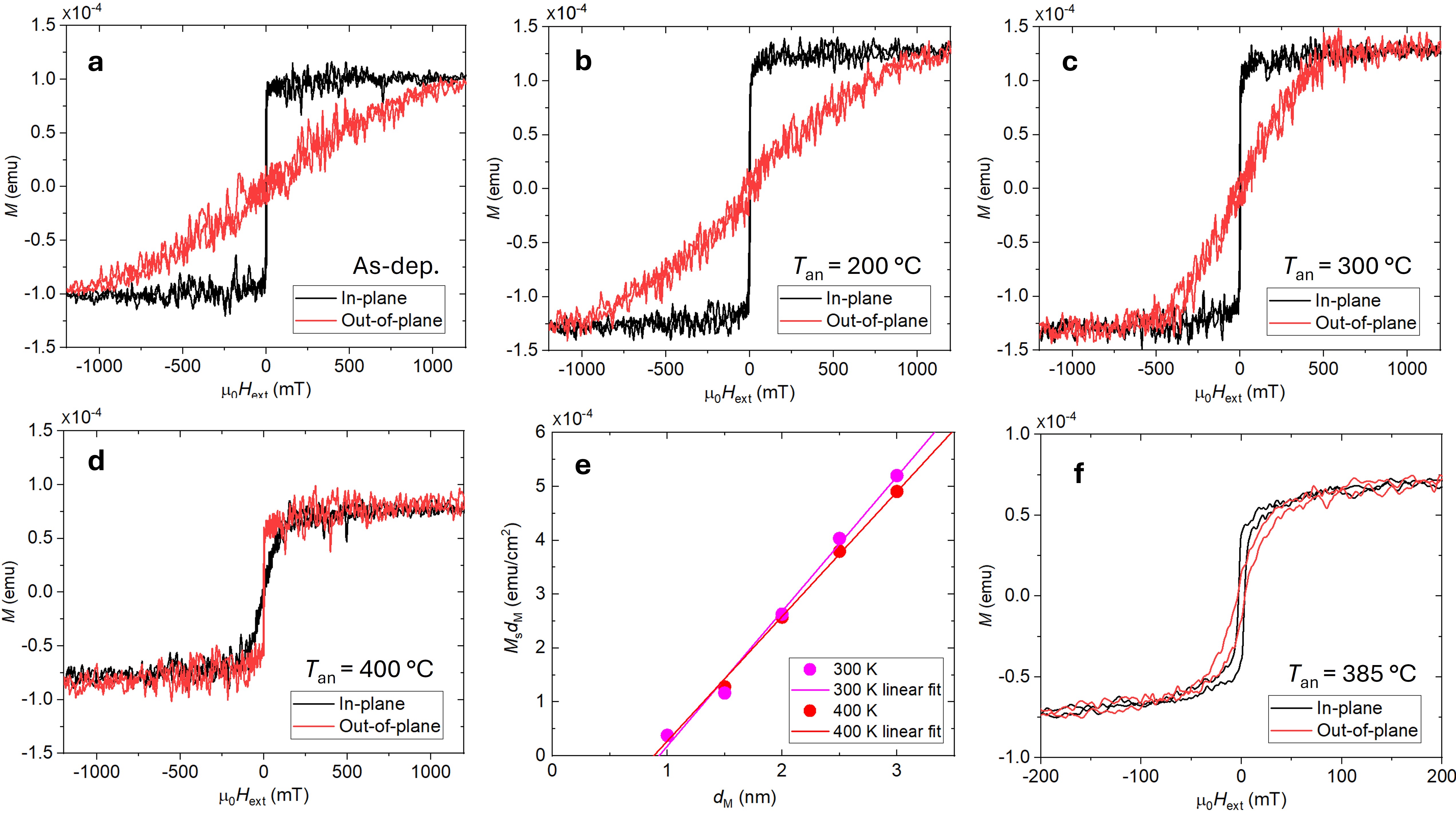}
  \caption{\textbf{a-d}, VSM characterisation of both in-plane and out-of-plane magnetisation-field ($M$-$H$) loops of a W\rev{\textbar}CoFeB (2 nm)\rev{\textbar}MgO (3 nm) stack for as-deposited (\textbf{a}) and subsequently annealed at different temperatures ($T_\text{an}$) of 200~$^\circ$C (\textbf{b}), 300~$^\circ$C (\textbf{c}) and 400~$^\circ$C (\textbf{d}). \textbf{e}, The product of saturation magnetisation ($M_\text{s}$) extracted by VSM and nominal thickness of CoFeB thickness ($d_\text{M}$) as a function of $d_\text{M}$ for the samples after annealed at 300 and 400~$^\circ$C. \textbf{f}, $M$-$H$ loops for another W\rev{\textbar}CoFeB (2 nm\rev{\textbar}MgO (3 nm) sample after annealing at 385~$^\circ$C.}
  \label{fig:MH}
\end{figure}

The magnetic dipole interaction generates an easy-plane magnetic anisotropy in thin films that is proportional to saturation magnetisation ($M_\text{s}$). It competes with the perpendicular anisotropy energy (\(K_{\perp}\)) originating from the spin-orbit interaction at the interface between CoFeB and MgO~\cite{Ikeda_NMater2010x,Yang_PRB2011x}. \(K_{\perp}\) can be tuned by various material parameters (e.g. crystallisation, chemical composition and morphology) we can control by post-annealing that only moderately affects $M_\text{s}$. Due to this difference in annealing dependence between \(K_{\perp}\) and $M_\text{s}$, we are able to find an annealing temperature where the effective magnetisation $M_\text{eff} = M_\text{s} - 2K_{\perp}/(\mu_0 M_\text{s}) \approx 0$ ($\mu_0$ is the magnetic permeability of the vacuum). We found that the annealing temperature to realise this condition slightly varies for different grown stacks for our study.

Figures~\ref{fig:MH}a-d show room-temperature $M$-$H$ loops of a W|CoFeB (2 nm)|MgO stack for different annealing temperatures ($T_\text{an}$). We performed this annealing in a vacuum furnace for one hour at $T_\text{an}$. We observe a transition from an easy-plane magnet to an out-of-plane easy-axis magnet when increasing $T_\text{an}$ between 300 $^\circ$C and 400 $^\circ$C. We fine-tuned the annealing temperature to minimise $M_\text{eff}$ for CoFeB thicknesses ($d_\text{M}$) and plot the product of $M_\text{s}$ and $d_\text{M}$ as a function of $d_\text{M}$ in Fig.~\ref{fig:MH}e. These results point to a magnetically dead layer of approximately 1 nm. Figure~\ref{fig:MH}f displays $M$-$H$ loops for another W|CoFeB (2 nm)|MgO (3 nm) sample for $T_\text{an} = 385$~$^\circ$C before we patterned this stack into transport devices that provide results for the main manuscript.

\newpage
\section*{\rev{Supplementary Note 2: Scanning transmission electron microscopy characterisation}}

\begin{figure}[bh]
    \centering
    \includegraphics[width=1.0\linewidth]{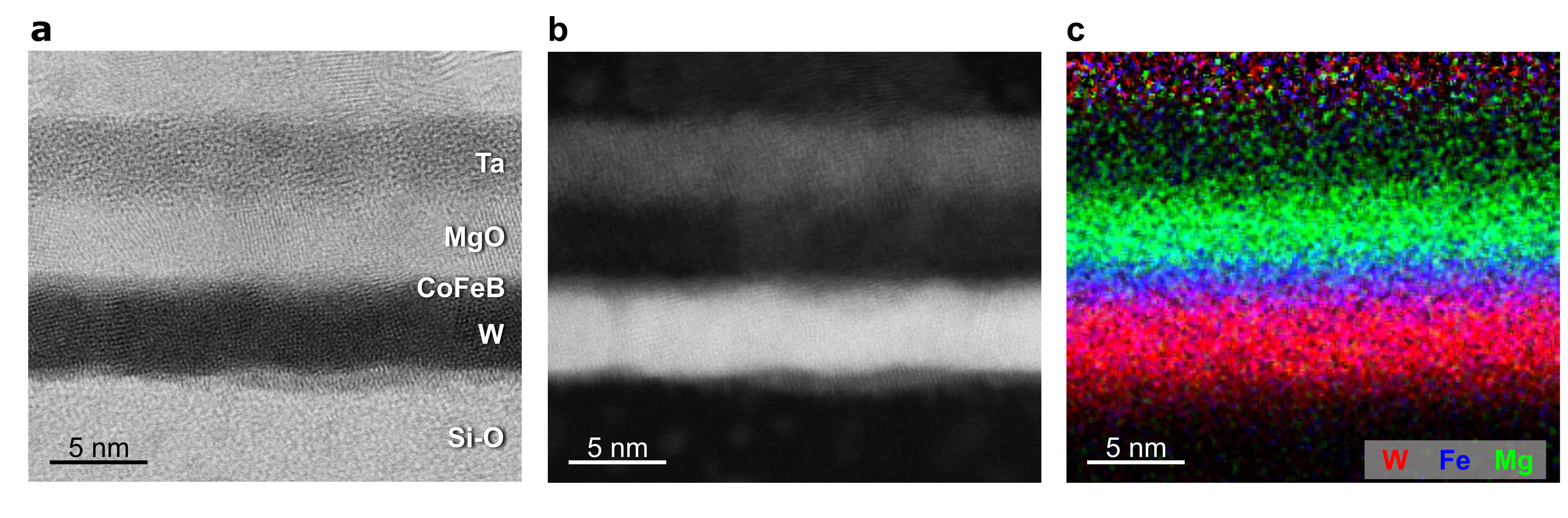}
  \caption{\rev{\textbf{a-b}, Cross-sectional BF (\textbf{a}) and HAADF (\textbf{b}) STEM images of the W\textbar CoFeB\textbar MgO stack. \textbf{c}, STEM-EDS elemental maps of the sample.}}
  \label{fig:TEM}
\end{figure}

\rev{The structural characterisation of the post-annealed W|CoFeB|MgO sample was carried out using scanning transmission electron microscopy (STEM). A cross-sectional STEM specimen was prepared from an Si substrate|Si-O|W|CoFeB|MgO stacks with a Ta capping layer using focused ion beam systems. The STEM observation was performed using a JEM-ARM200F (JEOL Ltd.) equipped with an energy dispersive x-ray spectroscopy (EDS) detector. Figures \ref{fig:TEM}a-b show the cross-sectional bright field (BF) and high-angle annular dark-field (HAADF) STEM images, respectively, where we confirm clear contrasts for W and MgO layers owing to their relatively large thicknesses. The CoFeB layer can also be recognised between the two layers where partial crystallisation of CoFeB due to the post-annealing is observed. EDS elemental mapping in Fig. \ref{fig:TEM}c further confirms the presence of Fe atoms, indicating that the clear layer stacking structure of W|CoFeB|MgO has been achieved. }

\newpage
\section*{Supplementary Note \rev{3}: magnetoresistance (MR)}
\begin{figure}[bh]
    \centering
    \includegraphics[width=1.0\linewidth]{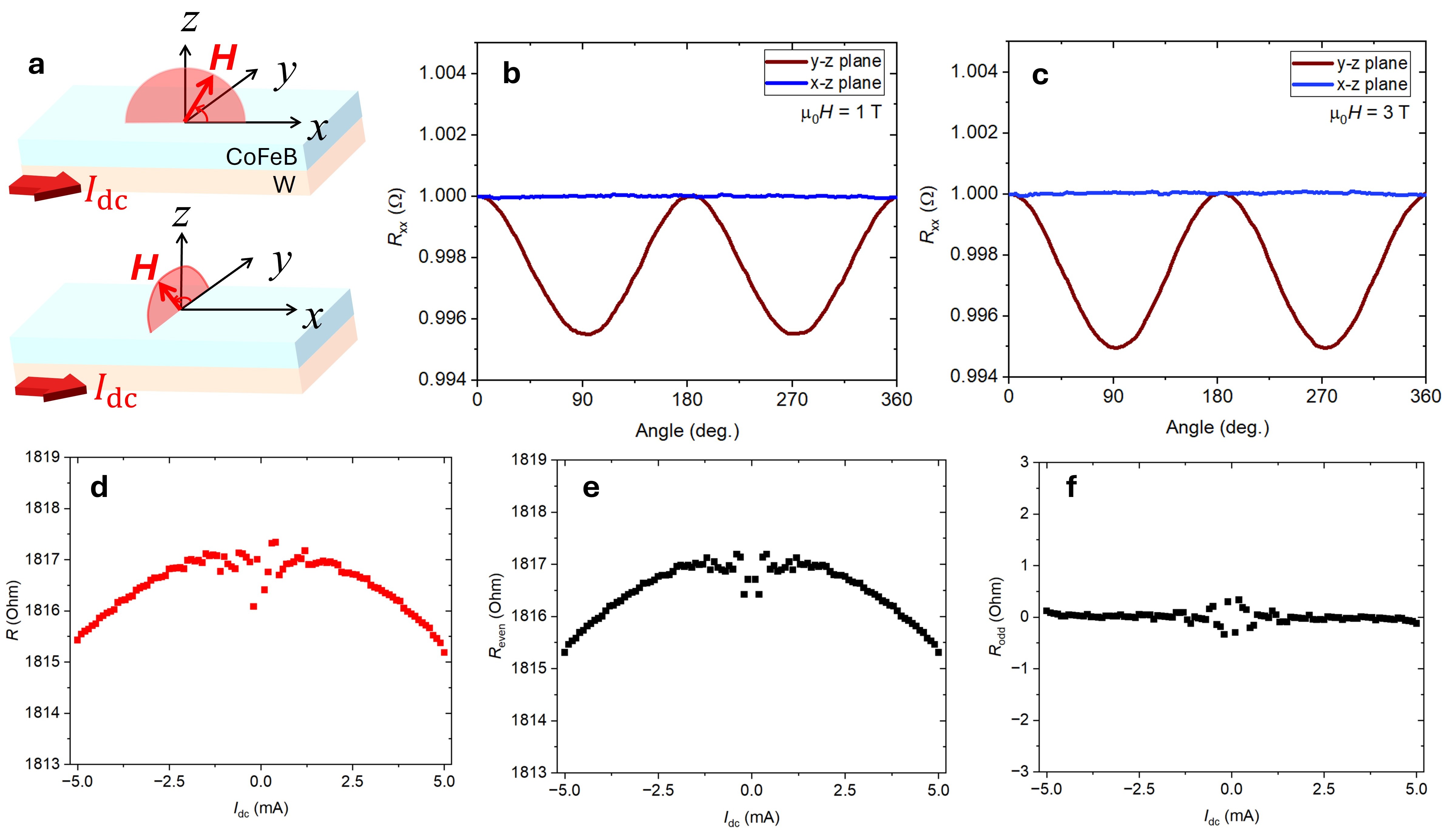}
  \caption{\textbf{a}, magnetoresistance configurations for $x$-$z$ (top) and $y$-$z$ (bottom) field rotations while a dc electric current ($I_\text{dc}$) flows along the $x$ direction for both cases. \textbf{b-c}, Fixed-field angular-dependent magnetoresistance measurements at the magnetic fields of 1 T (\textbf{b}) and 3 T (\textbf{c}) respectively, and performed at room temperature. $\textbf{d}$, Resistance ($R$) measurements by sweeping dc currents ($I_\text{dc}$) while applying a constant magnetic field at 516 mT. Artifact due to a small residual current from the source meter has been extracted before plotting $R$. $\textbf{e(f)}$, the even(odd) component $R_\text{even}$($R_\text{odd}$) of $R$ with respect to $I_\text{dc}$ reversal. The scale of the resistance change is fixed for 6 Ohms to compare the magnitude of the resistance change directly side by side.}
  \label{fig:smr}
\end{figure}

We measured the room temperature MR of MgO(3~nm)|CoFeB(2~nm)|W(3~nm) stacks as shown in Fig.~\ref{fig:smr}. Because both CoFeB and W layers are metallic, the anisotropic magnetoresistance (AMR)~\cite{Ritzinger_RSOpen2023x} in CoFeB coexists with spin-Hall magnetoresistance (SMR)~\cite{Nakayama_PRL2013x,Chen_PRB2013x} in W. Although these mechanisms cannot be distinguished when rotating the magnetisation in the film plane (e.g. those in Fig. 3 of the main manuscript), two high-symmetry out-of-plane rotations (Fig.~\ref{fig:smr}a) do lead to different angular dependencies to quantify individual components experimentally. The AMR depends on the angle between the magnetisation and the current directions (the $x$ direction in the Fig.~\ref{fig:smr}a schematics), whereas the SMR senses the angle between magnetisation and the polarisation of the spin Hall current (in our case along the $y$ direction). In Fig.~\ref{fig:smr}(c), the MR in the $y$-$z$ plane for a constant field of 1 T(3 T) oscillates, while the MR in  the $x$-$z$ scan is negligibly small, i.e. the SMR dominates. Because of this, we rule out the current-induced self-torque in the CoFeB layer from our analysis.

\rev{Next, we discuss the relationship between our measurements presented in Fig.~3 in the main text and the unidirectional magnetoresistance (UMR)\cite{Avci_PRL2018x}, or more specifically unidirectional SMR\cite{Avci_NPhys2015x}. Among different mechanisms of UMR\cite{Avci_NPhys2015,Avci_PRL2018,Borisenko_APL2018,Liu_PRL2021}, the magnon contribution is responsible for the leading order $I_{\rm dc}$ dependence of $\langle M_y^2 \rangle $. When the magnetisation dynamics has large amplitudes, $\langle M_y^2 \rangle $ as a function of $I_{\rm dc}$ is highly non-linear and contains both odd and even components, see Figs.~3a-b in the main text. To assess UMR components that are unrelated to magnons, we measured the resistance at a high magnetic field of 516 mT that strongly suppresses the current-induced modulation of $\langle M_{y}^2 \rangle$.  In Fig.~\ref{fig:smr}d we plot the resistance $R$ (after subtracting an instrumental artifact caused by a finite residual current in the source meter).  The UMR contribution to $R$ is odd under current reversal and captured by $R_\text{odd}(I_\text{dc})=(R(I_\text{dc})-R(-I_\text{dc}))/2$. However, Fig.~\ref{fig:smr}e-f shows an $I_\text{dc}$ dependence of $R$ that is dominated by the even component $R_\text{even}(I_\text{dc})=(R(I_\text{dc})+R(-I_\text{dc}))/2$, which signals Joule heating. Repeating these measurements at 618, 718, 816, 902 and 978 mT (data not shown) confirm that a non-magnonic contribution to the UMR is negligibly small.  }

\newpage
\section*{Supplementary Note \rev{4}: Theoretical Modelling of ST-FMR under Dynamical Stabilisation}
\begin{figure}[hb]
    \centering
    \includegraphics[width=0.8\linewidth]{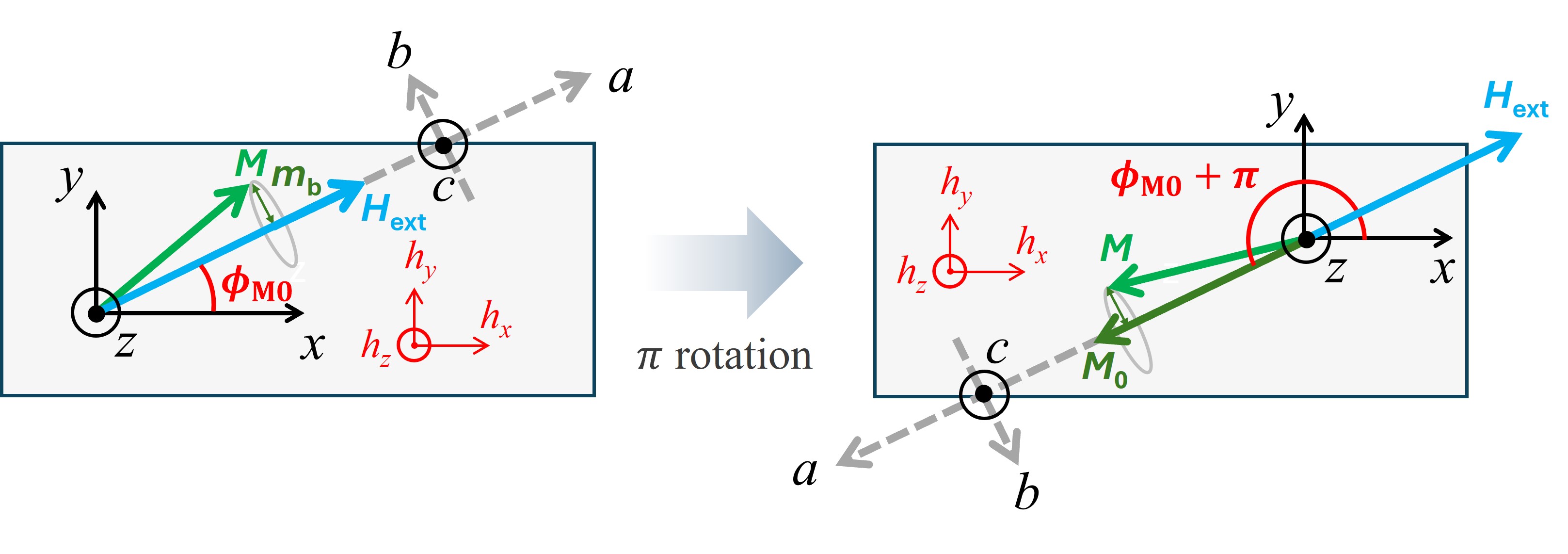}
  \caption{The coordinate systems used in our study are shown and both schematics are top-views of our devices. For our analysis of magnetisation dynamics, we use the $a$-$b$-$c$ coordinates defined by the direction of magnetic field, in other words, the direction of the time-averaged moment for the $a$ axis. Due to this definition, this coordinate system depends on the orientation of magnetic field and here we show the case of magnetisation reversal where the $b$ axis has been flipped.}
  \label{fig:coordinate}
\end{figure}

We start from Landau-Lifshitz-Gilbert equation for the macrospin augmented by the SOT~\cite{Liu_PRL2011x}
\begin{equation}
    \frac{d\bm{M}}{dt} = -\gamma \mu _0 \bm{M} \times \left( \bm{H}_{\rm ext} -M_{\rm eff}\frac{M_{z}}{M_{\rm s}}\hat{\bm{z}} + \bm{h} \right) + \frac{\alpha }{M_{\rm s}} \bm{M} \times \frac{d\bm{M}}{dt} -  \gamma \mu _0 \beta \left( I_{\rm dc}+I_{\rm mw} \cos \omega t \right) \frac{\bm{M}}{M_{\rm s}} \times \left( \bm{M} \times \hat{\bm{y}}\right) ,\label{eq:LLG}
\end{equation}
with symbols defined in the main text, except for $\omega =2\pi f$ and the rf Oersted field $\bm{h}=h\hat{\bm{y}}$ generated by the charge current~\cite{Liu_PRL2011x}. We disregard here the spin waves with nonzero wavelengths and thermal fluctuations. If the SOT is caused by the injection of the spin current generated by the spin-Hall effect, $\beta = \hbar \theta _{\rm SH}/2e\mu _0 M_{\rm s} wd_{\rm M} d_{\rm W}$ where $\hbar , \theta _{\rm SH}, e , w, d_{\rm M} ,d_{\rm W}$ are the reduced Planck constant, spin-Hall angle of W, elementary electric charge, width of the multilayer stack, and the thicknesses of the CoFeB and W layers respectively. In our setup, the W\rev{\textbar}CoFeB interface may well play a crucial role in deciding the magnitude of SOT, so that we choose to use the same symbol $\beta $ to avoid specifying the microscopic mechanism of SOT. We write $\bm{M} = \bm{M}_0 + \delta \bm{m} $ where $\bm{M}_0$ is independent of time, and treat $\delta \bm{m}$ as a small perturbation. Denoting the Fourier transform in time by a tilde, the linearized LLG equation reads
\begin{equation}
    \left( i\omega -\kappa + \gamma \mu _0 \beta I_{\rm dc} \frac{M_{0y}}{M_{\rm s}} \right) \widetilde{\delta \bm{m}} + \left( \gamma \mu _0 \bm{H}_{\rm ext} - i\alpha \omega \frac{\bm{M}_0}{M_{\rm s}} \right) \times \widetilde{\delta \bm{m}} = \gamma \mu _0 \frac{\bm{M_0}}{M_{\rm s}} \times \widetilde{\bm{h}} + \gamma \mu _0 \beta \widetilde{I}_{\rm mw} \frac{\bm{M}_0}{M_{\rm s}} \times \left( \frac{\bm{M}_0}{M_{\rm s}} \times \hat{\bm{y}} \right) , \label{eq:linear1}
\end{equation}
where $\widetilde{I}_{\rm mw}= I_{\rm mw}/2$ and $\kappa >0$ is a phenomenological relaxation rate on top of the Gilbert damping that may depend on the reference state $\bm{M}_0$ around which the linearisation is carried out. $\kappa >0$ acts as an additional stabilisation of the reference state. The atomistic simulations did not include $\kappa $, which is therefore not essential to achieve the dynamical stabilization of the anti-parallel state. However, we keep it here because the fitting of experimental data indicates the presence of additional dissipation processes beyond the Gilbert damping. Introducing polar coordinates of the Bloch sphere
\begin{equation}
    \frac{\bm{M}_0}{M_{\rm s}} = \begin{pmatrix}
        \sin \theta _M \cos \phi _M \\
        \sin \theta _M \sin \phi _M \\
        \cos \theta _M \
    \end{pmatrix}, \label{eq:polar_angles}
\end{equation}
we define a ground-state-adapted Cartesian frame $\{ \bm{a},\bm{b},\bm{c} \}$ (illustrated in Fig.~\ref{fig:coordinate} for $\theta _M =\pi /2$) as
\begin{equation}
    \bm{a} = \frac{\bm{M}_0}{M_{\rm s}} , \quad \bm{b} = \begin{pmatrix}
        -\sin \phi _M \\
        \cos \phi _M \\
        0 \\
    \end{pmatrix}, \quad \bm{c} = \bm{a} \times \bm{b} = \begin{pmatrix}
        -\cos \theta _M \cos \phi _M \\
        -\cos \theta _M \sin \phi _M \\
        \sin \theta _M \\
    \end{pmatrix} .
\end{equation}
With $H_{\rm ext}^a = \bm{H}_{\rm ext}\cdot \bm{a}, H_{\rm ext}^b = \bm{H}_{\rm ext}\cdot \bm{b}$, etc., Eq.~(\ref{eq:linear1}) becomes
\begin{equation}
    \begin{pmatrix}
        H^a_{\rm ext} -M_{\rm eff} \cos ^2 \theta _M - \frac{i\alpha \omega }{\gamma \mu _0} & \frac{i\omega -\kappa }{\gamma \mu _0} -\beta I_{\rm dc} \sin \theta _M \sin \phi _M \\
        \frac{-i\omega +\kappa }{\gamma \mu _0} + \beta I_{\rm dc}\sin \theta _M \sin \phi _M & H^a_{\rm ext} -M_{\rm eff}\cos 2\theta _M -\frac{i\alpha \omega }{\gamma \mu _0} \\
    \end{pmatrix} \begin{pmatrix}
        \widetilde{\delta m^b} \\
        \widetilde{\delta m^c} \\
        \end{pmatrix} = \begin{pmatrix}
            \widetilde{h} \cos \phi _M + \beta  \widetilde{I_{\rm mw}}\cos \theta _M \sin \phi _M \\
            -\widetilde{h}\cos \theta _M \sin \phi _M +\beta \widetilde{I_{\rm mw}}\cos \phi _M \\
        \end{pmatrix} . \label{eq:linear2}
\end{equation}
To the leading order terms in $\alpha $, $\kappa /\omega $, and $\beta $
\begin{eqnarray}
\widetilde{\delta m^b} =& \frac{ \left\{ \left( H^a_{\rm ext}-M_{\rm eff}\cos 2\theta _M \right) \cos \phi _M +\frac{i\omega }{\gamma \mu _0}\cos \theta _M \sin \phi _M \right\} \widetilde{h} +\left\{ \left( H^a_{\rm ext} -M_{\rm eff}\cos 2\theta _M \right) \cos \theta _M \sin \phi _M - \frac{i\omega }{\gamma \mu _0}\cos \phi _M \right\}\beta \widetilde{I_{\rm mw}}}{\left( H^a_{\rm ext}- H^+_{\rm res} \right) \left( H^a_{\rm ext} + H^-_{\rm res} \right) -\frac{i\omega }{\gamma \mu _0}\left\{ \alpha \left( 2H^a_{\rm ext} - H_{\rm res}^+ +H_{\rm res}^- \right) +\frac{2\kappa }{\gamma \mu _0} + \beta I_{\rm dc}\sin \theta _M \sin \phi _M \right\} } , \\
\widetilde{\delta m^c} =& \frac{ -\left\{ \left( H^a_{\rm ext}-M_{\rm eff}\cos ^2\theta _M \right) \cos \theta _M \sin \phi _M  - \frac{i\omega }{\gamma \mu _0}\cos \phi _M \right\} \widetilde{h} +\left\{ \left( H^a_{\rm ext} -M_{\rm eff}\cos ^2\theta _M \right) \cos  \phi _M + \frac{i\omega }{\gamma \mu _0}\cos \theta _M \sin \phi _M \right\}\beta \widetilde{I_{\rm mw}} }{\left( H^a_{\rm ext}- H^+_{\rm res} \right) \left( H^a_{\rm ext} + H^-_{\rm res} \right) -\frac{i\omega }{\gamma \mu _0}\left\{ \alpha \left( 2H^a_{\rm ext} - H_{\rm res}^+ + H_{\rm res}^-  \right) +\frac{2\kappa }{\gamma \mu _0} + \beta I_{\rm dc}\sin \theta _M \sin \phi _M \right\} } ,
\end{eqnarray}
where we introduced the provisional resonance fields $H^{\pm }_{\rm res}$ for the thermal equilibrium (+) and dynamically stabilised (-) state
\begin{equation}
H_{\rm res}^{\pm }=  \sqrt{\frac{\omega ^2}{\gamma ^2 \mu _0^2} + \frac{M_{\rm eff}^2 \sin ^4 \theta _M}{4}} \pm \frac{M_{\rm eff}\left( 3\cos ^2 \theta _M -1\right)}{2}   .
\end{equation}

The SMR under applied currents $I_{\rm dc}, I_{\rm mw}\cos \omega t$ generates a dc voltage
\begin{equation}
    V_{\rm dc} = \left\{ R_0 + \Delta R_{\rm SMR} \left( 1- \sin ^2 \theta _M \sin ^2 \phi _M \right) \right\} I_{\rm dc} -\frac{2\Delta R_{\rm SMR}}{M_{\rm s}} I_{\rm mw}\Re \left[ \widetilde{\delta m_y} \right] \sin \theta _M \sin \phi _M .
\end{equation}
With $\delta m_{y} = \delta m^b \cos \phi _M - \delta m^c \cos \theta _M \sin \phi _M$,
\begin{align}
\Re \left[ \widetilde{\delta m_{y}} \right] =&  \Re \left[ \frac{ \left( H^a_{\rm ext} -M_{\rm eff}\cos 2\theta _M \right) \cos ^2 \phi _M + \left( H^a_{\rm ext} -M_{\rm eff}\cos ^2 \theta _M \right) \cos ^2 \theta _M \sin ^2 \phi _M  }{\left( H^a_{\rm ext} -H^+_{\rm res} \right) \left( H^a_{\rm ext} +H^-_{\rm res} \right)-\frac{i\omega }{\gamma \mu _0} \left\{ \alpha \left( 2H_{\rm ext}^a -H_{\rm res}^+ +H_{\rm res}^- \right) + \frac{2\kappa }{\gamma \mu _0} + \beta I_{\rm dc}\sin \theta _M \sin \phi _M \right\} } \right] \widetilde{h} \label{eq:ST-FMR_general}  \\
&+ \Im \left[ \frac{\frac{\omega }{\gamma \mu _0}\left( \cos ^2 \phi _M + \cos ^2 \theta _M \sin ^2 \phi _M \right) +iM_{\rm eff}\cos \theta _M \sin ^2 \theta _M \cos \phi _M \sin \phi _M }{\left( H^a_{\rm ext} -H^+_{\rm res} \right) \left( H^a_{\rm ext} +H^-_{\rm res} \right)-\frac{i\omega }{\gamma \mu _0} \left\{ \alpha \left( 2H_{\rm ext}^a -H_{\rm res}^+ +H_{\rm res}^- \right) + \frac{2\kappa }{\gamma \mu _0} + \beta I_{\rm dc}\sin \theta _M \sin \phi _M \right\} } \right] \beta \widetilde{I_{\rm mw}}  ,\nonumber
\end{align}
where $\tilde{h}$ is real since $h$ is in-phase with $I_{\rm mw}\cos \omega t$. This formula holds for arbitrary direction of $\bm{M}_0$, which does not necessarily behave like a simple Lorentzian.
%It can be approximated by a simple Lorentzian around $H_{\rm ext}^a = \pm H_{\rm res}^{\pm }$ only if $H^a_{\rm ext}, H_{\rm res}^{\pm }, \theta _M ,\phi _M$ are all slowly varying functions of $H_{\rm ext}$, which might not be satisfied in the presence of an out-of-plane component of $\bm{H}_{\rm ext}$.

Next, we focus on $M_{\rm eff}>0$ and $\bm{M}_0 \parallel \pm \bm{H}_{\rm ext}$ \emph{i.e.}, $\theta _M = \pi /2$ and $\phi _M = \phi $ or $\phi _M = \phi +\pi$. The stability of these reference states will be studied in the next section. $\bm{M}_0 \parallel \pm \bm{H}_{\rm ext}$ implies $H^a_{\rm ext} = \pm H_{\rm ext}$, and different resonance fields $H^{\pm }_{\rm res}$ for the parallel and anti-parallel states respectively. Assuming constant $\theta _M ,\phi _M$, one can replace $H_{\rm ext}^a $ in Eq.~(\ref{eq:ST-FMR_general}) by $\pm H^{\pm }_{\rm res}$ except in the terms rapidly approaching zero in the denominators close to the resonance. We recover the Lorentzian dependence on $H_{\rm ext}$ when the imaginary part of the denominator is constant. However, we tune $I_{\rm dc}$ such that the imaginary part vanishes at a field near the resonance. The experiment indeed shows deviations from the Lorentzian lineshape, which is qualitatively consistent with Eq.~(\ref{eq:ST-FMR_general}). In the following, we disregard this complication for simplicity by using the expression $V_{\rm dc} = \left\{ R_0 + \Delta R_{\rm SMR}\left( 1- \sin ^2 \phi \right) \right\}  I_{\rm dc} + V_{\rm sym} + V_{\rm asy}$ with $V_{\rm sym},V_{\rm asy}$ the symmetric and anti-symmetric parts of the ST-FMR signal given by
\begin{eqnarray}
    V_{\rm sym} &=& -2\Delta R_{\rm SMR}\beta \widetilde{I_{\rm mw}}^2 \sqrt{\frac{1}{1+\gamma ^2 \mu _0^2 M_{\rm eff}^2/4\omega ^2}} \frac{\pm \Delta H_{\pm }\cos ^2 \phi \sin \phi }{\left( H_{\rm ext} -H^{\pm }_{\rm res} \right) ^2 + \Delta H_{\pm }^2} , \label{eq:Vsym} \\
    V_{\rm asy} &=& -4\Delta R_{\rm SMR}\widetilde{I_{\rm mw}} \widetilde{h} \frac{ H^{\pm }_{\rm res} \pm M_{\rm eff}}{H_{\rm res}^+ +H_{\rm res}^-}  \frac{\left( H_{\rm ext}-H_{\rm res}^{\pm } \right) \cos ^2 \phi \sin \phi }{\left( H_{\rm ext} -H_{\rm res}^{\pm } \right) ^2 + \Delta H_{\pm }^2} ,\label{eq:Vasy}
\end{eqnarray}
where the linewidths for the parallel and anti-parallel reference states read
\begin{equation}
    \Delta H_{\pm } = \Delta H_0 \pm \frac{\alpha \omega }{\gamma \mu _0} \pm \sqrt{\frac{1}{1+\gamma ^2 \mu _0^2 M_{\rm eff}^2/4\omega ^2}}\beta I_{\rm dc} \sin \phi , \quad \Delta H_0 = \frac{\kappa }{\gamma \mu _0} \sqrt{\frac{1}{1+\gamma ^2 \mu _0^2 M_{\rm eff}^2 /4\omega ^2}}. \label{eq:linewidth1}
\end{equation}
Although we used the notation $\Delta H_0$ to suggest it corresponds to the inhomogeneous broadening, $\kappa $ was not introduced as such and we regard both $\Delta H_0$ and $\kappa$ as pure fitting parameters without any association to underlying physics. The FMR experiments are well represented by a symmetric Lorentzian (\textit{i.e.}, $V_\text{sym} \gg V_\text{asy}$), which indicates that field-like torques are very small \textit{i.e.}, $\beta \left| I_{\rm mw} \right| \gg \left| \bm{h} \right|$. For fixed $\phi $, the angle for the magnetic field direction, we obtain a familiar expression of linewidth in the parallel state $\Delta H_+$ in which $I_{\rm dc}\sin \phi <0 $ counteracts the Gilbert damping. The linewidth of the anti-parallel state $\Delta H_- $ is negative when SOT and $\Delta H_0$ vanish, reflecting that it is unstable at a maximum of the free energy. A large negative $I_{\rm dc}\sin \phi$ turns $\Delta H_- $ positive, which indicates stability of the anti-parallel state that can be detected by a sign change of $V_{\rm sym}$. In the next section, we will show that for a sufficiently large negative $I_{\rm dc}\sin \phi $, $\Delta H_- >0$ for $H_{\rm ext}$ near $H_{\rm res}^-$ so that the sign of $V_{\rm sym}$ indicates whether $\bm{M}_0$ is parallel or anti-parallel to $\bm{H}_{\rm ext}$. Substituting $\omega = 2\pi f$ into Eqs.~(\ref{eq:Vsym}) and (\ref{eq:linewidth1}) yields Eqs.~(2) and (3) in the main text respectively.

\newpage
\section*{Supplementary Note \rev{5}: Stability Diagram}

In this section, we describe how the bifurcation diagram Fig.~4a in the main text is generated. Namely, we analytically determine the conditions for stability of various time-independent solutions of Eq.~(\ref{eq:LLG}), referred to as fixed points, for $\bm{H}_{\rm ext} = H_{\rm ext}\hat{\bm{y}}$ and $h=I_{\rm mw}=0$. For $\phi \neq \pm \pi /2 $, the stability conditions should be well approximated by replacing $I_{\rm dc}$ in the formulae for $\phi =\pm \pi /2$ by $I_{\rm dc}\sin \phi$ with the deviation proportional to $\beta I_{\rm dc}/\left( H_{\rm ext}\pm M_{\rm eff}\right)$ that we assume to be small. We first enumerate all the possible fixed points in the polar coordinates Eq.~(\ref{eq:polar_angles}) by setting $d\bm{M}/dt=0$ in Eq.~(\ref{eq:LLG}), which leads to
\begin{equation}
\begin{pmatrix}
    1 & \eta _{\rm STT}\cos \theta _M \\
    -\eta _{\rm STT} & \cos \theta _M \\
\end{pmatrix} \begin{pmatrix}
    \cos \phi _M \\
    \sin \phi _M \\
    \end{pmatrix} = -\nu\begin{pmatrix}
        0 \\
        \cos \theta _M \sin \theta _M \\
    \end{pmatrix} , 
\end{equation}
where $\eta _{\rm STT} =\beta I_{\rm dc}/H_{\rm ext}$ is the dimensionless STT efficiency and $\nu = M_{\rm eff}/H_{\rm ext}$ the anisotropy. The solutions with $\cos \theta _M =0$, \textit{i.e.}, magnetisation in the plane, are the parallel and anti-parallel states $\phi _M = \pm \pi /2$. For $\cos \theta _M \neq 0$, the canting angle $\theta _M$ solves the equation
\begin{equation}
    \eta _{\rm STT}^2 \sin ^4 \theta _M - \left( 1+\eta _{\rm STT}^2 \right) \sin ^2 \theta _M + \frac{1}{\nu ^2}\left( 1+\eta _{\rm STT}^2 \right) ^2 = 0.
\end{equation}
We restrict ourselves to relatively small currents with $-1<\eta _{\rm STT}<1$. A solution, which is unique up to the sign, exists when $\left| \nu \right| > 1+\eta _{\rm STT}^2$;
\begin{equation}
    \sin ^2 \theta _M = \frac{1+\eta _{\rm STT}^2}{2\eta _{\rm STT}^2}\left( 1-\sqrt{1-\frac{4\eta _{\rm STT}^2}{
\nu ^2}
    } \right) , \quad \begin{pmatrix}
        \cos \phi _M \\
        \sin \phi _M \\
    \end{pmatrix} = -\frac{\nu \sin \theta _M}{1+\eta _{\rm STT}^2}\begin{pmatrix}
        -\eta _{\rm STT}\cos \theta _M \\
        1 \\
    \end{pmatrix} . \label{eq:out-of-plane}
\end{equation}
For $M_{\rm eff}<0$, we recover the solution at thermal equilibrium in which the in-plane $H_{\rm ext}$ and the PMA compete. To summarise, the fixed points are: (i) the thermal equilibrium state $\theta _M =\pi /2 ,\phi _M = \pi /2$, (ii) the inverted state $\theta _M = \pi /2, \phi _M =-\pi /2$, and (iii) the pair of out-of-plane states given in Eq.~(\ref{eq:out-of-plane}) that exist only for a sufficiently large $\left| M_{\rm eff} \right| $. Thier stability follows from Eq.~(\ref{eq:linear2}) with right-hand-sides set to zero, which leads to complex eigenfrequencies
\begin{eqnarray}
\frac{\left( 1+\alpha ^2 \right) \omega }{\gamma \mu _0 H_{\rm ext}} &=& -i \left\{\alpha \left( \sin \theta _M \sin \phi _M -\nu \frac{ 3\cos ^2 \theta _M -1}{2} \right)+ \frac{\kappa }{\gamma \mu _0 H_{\rm ext}}+\eta _{\rm STT} \sin \theta _M \sin \phi _M \right\} \nonumber \\
&& \pm \Bigg[  \left\{ \sin \theta _M \sin \phi _M -\nu \cos ^2 \theta _M -\alpha \left( \frac{\kappa }{\gamma \mu _0 H_{\rm ext}} +\eta _{\rm STT}\sin \theta _M \sin \phi _M \right) \right\} \nonumber \\
&& \times \left\{ \sin \theta _M \sin \phi _M - \nu \cos 2\theta _M  -\alpha \left( \frac{\kappa }{\gamma \mu _0 H_{\rm ext}}+\eta _{\rm STT} \sin \theta _M \sin \phi _M \right) \right\} -\frac{\alpha ^2 \nu ^2 \sin ^4 \theta _M}{4} \Bigg]^{1/2} . \label{eq:eigenfrequencies}
\end{eqnarray}
The fixed point is stable if and only if the imaginary parts of both frequencies are negative.

\subsection*{1. Parallel state}
The eigenfrequencies take the familiar form
\begin{eqnarray}
    \frac{\left( 1+\alpha ^2 \right) \omega }{\gamma \mu _0 H_{\rm ext}} &=& -i\left\{ \alpha \left( 1 +\frac{\nu }{2} \right) + \frac{\kappa }{\gamma \mu _0 H_{\rm ext}} + \eta _{\rm STT} \right\} \nonumber \\
    && \pm \sqrt{ \left\{ 1 -\alpha \left( \frac{\kappa }{\gamma \mu _0 H_{\rm ext}}+\eta _{\rm STT} \right) \right\} \left\{ 1 +\nu  -\alpha \left( \frac{\kappa }{\gamma \mu _0 H_{\rm ext}} +\eta _{\rm STT} \right) \right\} - \frac{\alpha ^2 \nu ^2}{4}} .
\end{eqnarray}
The fixed point is stable if (i) the imaginary part of the first term on the right-hand-side is negative
\begin{equation}
    -\eta _{\rm STT}  < \alpha \left( 1 +\frac{\nu }{2} \right) + \frac{\kappa }{\gamma \mu _0 H_{\rm ext}}, \label{eq:stability_parallel}
\end{equation} 
and (ii) the modulus of the imaginary part of the square root is smaller than that of the first term, \textit{i.e.},
\begin{equation}
     -\left( 1+\alpha ^2 \right) \left\{ 1+\nu  + \left( \frac{\kappa }{\gamma \mu _0 H_{\rm ext}} +\eta _{\rm STT}\right) ^2 \right\} <0.
\end{equation}
Both conditions are satisfied for an easy-plane anisotropy $M_{\rm eff}>0$ and a sufficiently small STT. The parallel state can be destabilised by a large negative STT or by an easy-axis anisotropy $M_{\rm eff} \lesssim -H_{\rm ext}$. Equation~(1) in the main text is derived by replacing "$<$" by "$=$" in Eq.~(\ref{eq:stability_parallel}) and multiplying both sides by $H_{\rm ext}$, under the identification $\Delta H_0^{\prime } = \kappa /(\gamma \mu _0 )$.

\subsection*{2. Anti-parallel state}

Substituting $\theta _M = \pi/2, \phi _M = -\pi /2 $ into Eq.~(\ref{eq:eigenfrequencies}) yields
\begin{eqnarray}
    \frac{\left( 1+\alpha ^2 \right) \omega }{\gamma \mu _0 H_{\rm ext}} &=& i\left\{ \alpha \left( 1 - \frac{\nu }{2} \right) - \frac{\kappa }{\gamma \mu _0 H_{\rm ext}} + \eta _{\rm STT} \right\} \nonumber \\
    && \pm \sqrt{ \left\{ 1 + \alpha \left( \frac{\kappa }{\gamma \mu _0 H_{\rm ext}}-\eta _{\rm STT} \right) \right\} \left\{ 1 - \nu  + \alpha \left( \frac{\kappa }{\gamma \mu _0 H_{\rm ext}} -\eta _{\rm STT} \right) \right\} - \frac{\alpha ^2 \nu ^2}{4}} .
\end{eqnarray}
As argued above, the stability conditions are
\begin{equation}
    -\eta _{\rm STT} > \alpha \left( 1-\frac{\nu }{2} \right) -\frac{\kappa }{\gamma \mu _0 H_{\rm ext}} ,  \label{eq:condition_ap_1} 
\end{equation}
and
\begin{equation}
    -\left( 1+\alpha ^2 \right) \left\{ 1-\nu  + \left( \frac{\kappa }{\gamma \mu _0 H_{\rm ext}} -\eta _{\rm STT} \right) ^2  \right\} <0 . \label{eq:condition_ap_2}
\end{equation}
A large $\kappa $ may by itself stabilise the state, which is simply how the phenomenological relaxation was introduced. We do not know whether this is a sensible prescription, but keep this parameter for the sake of uniformity across different fixed points. In practice, it becomes important only for fitting the experimental data with the results concerning the parallel fixed point. When $\alpha H_{\rm ext} > \kappa /\gamma \mu _0$, the condition (\ref{eq:condition_ap_1}) can be satisfied either for large negative $\eta _{\rm STT} $ or large positive $M_{\rm eff}$. Since the latter violates the second condition (\ref{eq:condition_ap_2}), $\eta _{\rm STT}<0$ is a necessary condition for the stability.

\subsection*{3. Out-of-plane states}

When using Eq.~(\ref{eq:out-of-plane}) in Eq.~(\ref{eq:eigenfrequencies})
\begin{eqnarray}
\frac{\left( 1+\alpha ^2 \right) \omega }{\gamma \mu _0 H_{\rm ext}} &=& i  \nu \left\{ \alpha - \frac{\alpha -2\eta _{\rm STT} +3\alpha \eta _{\rm STT}^2}{4\eta _{\rm STT}^2}\left( 1-\sqrt{1-\frac{4\eta _{\rm STT}^2}{\nu ^2}} \right) -\frac{\kappa }{\gamma \mu _0 M_{\rm eff}}  \right\} \nonumber \\
&& \pm \nu\Bigg[ \left\{  1- \frac{\eta _{\rm STT}\left( \alpha + \eta _{\rm STT}\right) }{2\eta _{\rm STT}^2}\left( 1-\sqrt{1-\frac{4\eta _{\rm STT}^2}{\nu ^2}}\right) +  \frac{\alpha \kappa }{\gamma \mu _0 M_{\rm eff}} \right\} \nonumber \\
&& \times \left\{ 1 - \frac{1 +\eta _{\rm STT}\left( \alpha +2\eta _{\rm STT}\right)}{2\eta _{\rm STT}^2} \left( 1-\sqrt{1-\frac{4\eta _{\rm STT}^2}{\nu ^2}}\right) +  \frac{\alpha \kappa }{\gamma \mu _0 M_{\rm eff}} \right\} -\frac{\alpha ^2 \sin ^4 \theta _M}{4} \Bigg] ^{1/2} .
\end{eqnarray}
The conditions for stability are (i) the imaginary part of the first term to be negative
\begin{equation}
\nu \left\{ \alpha -\frac{\alpha -2\eta _{\rm STT}+3\alpha \eta _{\rm STT}^2}{4\eta _{\rm STT}^2} \left( 1-\sqrt{1-\frac{4\eta _{\rm STT}^2}{\nu ^2} } \right) \right\} < \frac{\kappa }{\gamma \mu _0 H_{\rm ext}}  \label{eq:stability_out-of-plane}
\end{equation}
and (ii) the modulus of the imaginary part of the square root to be smaller than that of the first line
\begin{equation}
    \left\{ 1+\frac{1}{2}\left( 1-\sqrt{1-\frac{4\eta _{\rm STT}^2}{\nu ^2} } \right) \right\} \left\{ 1 -\frac{1-\eta _{\rm STT}^2}{2\eta _{\rm STT}^2}\left( 1-\sqrt{1-\frac{4\eta _{\rm STT}^2}{\nu ^2} } \right) \right\} + \left( \frac{\kappa }{\gamma \mu _0 M_{\rm eff}} -\frac{1}{2\eta _{\rm STT}}\left( 1-\sqrt{1-\frac{4\eta _{\rm STT}^2}{\nu ^2} } \right) \right) ^2 > 0  .
\end{equation}
The latter condition is satisfied whenever the fixed points exist since 
\begin{equation}
    \frac{1-\eta _{\rm STT}^2}{2\eta _{\rm STT}^2} \left( 1-\sqrt{1-\frac{4\eta _{\rm STT}^2}{\nu ^2} } \right) < \sin ^2 \theta _M \leq 1.
\end{equation}
When $\eta _{\rm STT}=0$, those fixed points are stable equilibria only for an easy-axis anisotropy $M_{\rm eff}<0$. For $\eta _{\rm STT}^2 <1$ and $\alpha \ll 1$, the STT reinforces or counteracts the Gilbert damping according to whether $\eta _{\rm STT} >0 $ or $\eta _{\rm STT}<0$. Therefore and counterintuitively, a large negative $\eta _{\rm STT}$ can stabilise out-of-plane states even for $M_{\rm eff}>0$ or destabilise them when $M_{\rm eff}<0$.

\subsection*{4. Bifurcation diagram}

The results of the previous subsections can be summarised in the space spanned by the dimensionless parameters $\eta _{\rm STT}$ and $\nu $. The separatrices, \textit{i.e.}, the curves across which the stability of at least one of the fixed points changes, read (upper inequality for stability):
\begin{description}
    \item[Parallel state] 
    \begin{eqnarray}
        \nu &\gtrless & -2 -\frac{2\eta _{\rm STT}}{\alpha } -\frac{2\kappa }{\gamma \mu _0 H_{\rm ext}} , \\
        \nu &\gtrless & -1 -\left( \eta _{\rm STT} + \frac{\kappa }{\gamma \mu _0 H_{\rm ext}}\right) ^2 . 
    \end{eqnarray}
    \item[Anti-parallel state]
    \begin{eqnarray}
        \nu & \gtrless & 2 + \frac{2\eta _{\rm STT}}{\alpha } -\frac{2\kappa }{\gamma \mu _0 H_{\rm ext}} , \\
        \nu &\lessgtr & 1 + \left( \eta _{\rm STT} -\frac{\kappa }{\gamma \mu _0 H_{\rm ext}} \right) ^2 . 
    \end{eqnarray}
    \item[Out-of-plane states]
    The region of existence for the fixed points is divided into two by the sign of $M_{\rm eff}$. For the easy-plane anisotropy $\nu >0$, 
    \begin{eqnarray}
        \nu & >& 1+\eta _{\rm STT}^2 , \\
        \nu & \lessgtr & \left( 1 + \frac{2\alpha \eta _{\rm STT}^2}{\alpha -2\eta _{\rm STT}+\alpha \eta _{\rm STT}^2} \right) \sqrt{\frac{\alpha -2\eta _{\rm STT}+\alpha \eta _{\rm STT}^2}{2\alpha }+\left( \frac{\kappa }{2\alpha \gamma \mu _0 H_{\rm ext}}\right) ^2}+ \frac{\kappa }{2\alpha \gamma \mu _0 H_{\rm ext}} \frac{\alpha -2\eta _{\rm STT} -\alpha \eta _{\rm STT}^2}{\alpha -2\eta _{\rm STT}+\alpha \eta _{\rm STT}^2}. \label{eq:separatrix_oop_1}
    \end{eqnarray}
    When the anisotropy is of easy-axis type $\nu <0$,
    \begin{eqnarray}
        \nu & <& -1-\eta _{\rm STT}^2 , \\
        \nu & \lessgtr & -\left( 1 + \frac{2\alpha \eta _{\rm STT}^2}{\alpha -2\eta _{\rm STT}+\alpha \eta _{\rm STT}^2} \right) \sqrt{\frac{\alpha -2\eta _{\rm STT}+\alpha \eta _{\rm STT}^2}{2\alpha }+\left( \frac{\kappa }{2\alpha \gamma \mu _0 H_{\rm ext}}\right) ^2}+ \frac{\kappa }{2\alpha \gamma \mu _0 H_{\rm ext}} \frac{\alpha -2\eta _{\rm STT} -\alpha \eta _{\rm STT}^2}{\alpha -2\eta _{\rm STT}+\alpha \eta _{\rm STT}^2}. \label{eq:separatrix_oop_2}
    \end{eqnarray}
    Note that for $\alpha \eta _{\rm STT}> 1-\sqrt{1-\alpha ^2 -\kappa ^2 /\gamma ^2 \mu _0^2 H_{\rm ext}^2}$, the square root becomes imaginary and the inequalities do not make sense. The stability can be determined directly from Eq.~(\ref{eq:stability_out-of-plane}), nevertheless; they are unstable for $\nu <0 $ and stable for $\nu >0$.
\end{description}
In specific ranges of $\eta _{\rm STT},\nu $, all the fixed points are unstable. The Poincaré-Bendixon's theorem excludes chaos or strange attractors as asymptotic states of the Bloch sphere. Therefore, in that region of parameter space, the attractor state should be a limit cycle referred to as auto-oscillation in magnetism. Although we are unable to exclude limit cycles as possible attractors when some of the fixed points are stable, numerical simulations show no an evidence of asymptotic behaviour other than convergence to a stable fixed point. This allows us to generate a bifurcation diagram that classifies the parameter regions according to their attractors. Figure~4a in the main text is an example for $\alpha =0.1, \kappa =0$.

\subsection*{5. Interpretation of the ST-FMR results}

The dynamical systems analysis suggests three regimes under our experimental condition $0<M_{\rm eff} \lesssim H_{\rm ext}$, \textit{viz.} $\bm{M}_0 \parallel \bm{H}_{\rm ext}$, $\bm{M}_0 \parallel -\bm{H}_{\rm ext}$, and a bistability between these two. Upon sweeping $I_{\rm dc} \sin \phi $ from positive to negative values, one should see the parallel, bistable, and anti-parallel states appear in this order. We measure ST-FMR instead at fixed $I_{\rm dc}$ and $\omega $ while sweeping $H_{\rm ext}$. This corresponds to following a straight line passing $\eta _{\rm STT}=\nu =0$ in the bifurcation diagram (Fig.~4a in the main text), during which the stability of the fixed points can change. The ST-FMR voltage expression given in Eq.~(\ref{eq:Vsym}) depends on the reference state, namely the upper and lower signs have to be used for the parallel and anti-parallel fixed points respectively. Therefore, within any magnetic field interval in which a stability change occurs, the usual fitting procedure of the voltage would not work. Indeed, at $f = \omega /2\pi = 2.5$ GHz in Fig.~2f in the main text, the absence of clear resonance can be interpreted as $H_{\rm res}$ at 2.5 GHz falling in that interval of stability change for $I_{\rm dc}=4$ mA and $\phi = 225^{\circ }$.

The $\pm $ sign in Eq.~(\ref{eq:Vsym}) suggests that the sign of the resonance peak in $V_{\rm dc}$ discriminates the parallel or anti-parallel fixed points. However, numerous complications exist. The assumption that $\Delta H_{\pm }$ does not change sign as a function of $\omega $ and $I_{\rm dc}$ does not hold according to Eq.~(\ref{eq:linewidth1}). The field linewidth at the critical current of a stability change of the fixed points at $H_{\rm ext} = H_{\rm res}^{\pm }$ reads
\begin{equation}
    \Delta H_{\pm } = \Delta H_0 \pm \frac{\alpha \omega }{\gamma \mu _0} \mp \left\{ \alpha \left( H_{\rm res}^{\pm }\pm \frac{M_{\rm eff}}{2} \right) \pm \frac{\kappa }{\gamma \mu _0} \right\} \sqrt{\frac{1}{1+\gamma ^2 \mu _0^2 M_{\rm eff}^2 /4\omega ^2}}=0. \label{eq:critical_width}
\end{equation}
This ensures that at least at the resonance field, the stability change of the fixed points corresponds to the change of sign of $\Delta H_{\rm res}^{\pm }$. A fixed point should be stable whenever a clear resonance around it is observable. Therefore, one may assume $\Delta H_{\pm }>0$ when fitting a resonance peak regardless of the reference state. If $\omega $ and $I_{\rm dc}$ were tuned to satisfy Eq.~(\ref{eq:critical_width}) exactly, the stability change between $H_{\rm ext}\lessgtr H_{\rm res}^{\pm }$ would prevent a clear Lorentzian signal. Moreover, as discussed in the previous section, $V_{\rm dc}$ is not Lorentzian fwhen $\Delta H_{\pm } \sim 0 $ for $H_{\rm ext} \sim H_{\rm res}^{\pm }$. Hence, the interpretation of an observed peak as a resonance according to Eq.~(\ref{eq:Vsym}) can be justified only if for the given $\omega $ and $I_{\rm dc}$ the stability of the fixed points does not change in the interval $H_{\rm res}^{\pm } -\Delta H_{\pm } \lesssim H_{\rm ext} \lesssim H_{\rm res}^{\pm }+\Delta H_{\pm } $. Only then, a sign change of the resonance peak would be evidence for a magnetisation reversal.

\newpage
\section*{Supplementary Note \rev{6}: Fokker-Planck Equation}

Thermal excitation induces random fluctuations of the magnetic order by a noise magnetic field $\bm{\xi }\left( t\right) $ (in units of T, \textit{i.e.}, magnetic flux density) in the LLG equation Eq.~(\ref{eq:LLG}). When terms of order $\alpha ^2$ are disregarded, the stochastic LLG equation in the absence of a microwave drive is equivalent to a Fokker-Planck equation for the probability distribution function $f\left( \bm{n},t\right) $ with $\bm{n}$ the dimensionless unit vector of the magnetisation~\cite{Garanin_1997x}.
\begin{equation}
    \frac{\partial f}{\partial t} = \gamma \mu _0 \frac{\partial }{\partial \bm{n}} \cdot \left[ \bm{n}\times \left\{ \bm{H}_{\rm ext}-M_{\rm eff}n_{z} \hat{\bm{z}} +\bm{n}\times \left( \alpha \bm{H}_{\rm ext} -\alpha M_{\rm eff}n_{z} \hat{\bm{z}} + \beta I_{\rm dc}\hat{\bm{y}} - \frac{\alpha k_B T}{\mu _0 M_{\rm s} V_{\rm a}}\frac{\partial }{\partial \bm{n}} \right) \right\} f  \right] .
\end{equation}
One can solve this equation in the relevant limit of vanishing magnetic anisotropy $M_{\rm eff}=0$ and an STT spin polarisation parallel to $\bm{H}_{\rm ext}$ \textit{i.e.}, $\hat{\bm{y}} \rightarrow \left( \hat{\bm{y}}\cdot \bm{H}_{\rm ext} \right) \bm{H}_{\rm ext}/H_{\rm ext}^2 = \bm{H}_{\rm ext}\sin \phi /H_{\rm ext}$. In the steady state $\partial f/\partial t=0$:
\begin{equation}
    f\left( \bm{n} \right) = \frac{\Delta _{\rm STT}}{4\pi \sinh \Delta _{\rm STT} } \exp \left( \Delta _{\rm STT} \frac{\bm{H}_{\rm ext}}{H_{\rm ext}}\cdot \bm{n} \right) ,
\end{equation}
where $\Delta _{\rm STT} $ is a thermal stability factor defined by
\begin{equation}
    \Delta _{\rm STT} = \frac{\mu _0 M_{\rm s} H_{\rm ext}V_{\rm a}}{k_B T} \left( 1+\frac{\beta I_{\rm dc}\sin \phi }{\alpha H_{\rm ext}} \right)  .
\end{equation}
The prefactor outside the curly brackets is the ratio between the Zeeman and thermal energies. $f$ reduces to the Boltzmann distribution in the absence of the STT. A current with $I_{\rm dc}\sin \phi < 0$ decreases $\Delta _{\rm STT}$ until the critical value $\beta I_{\rm dc}\sin \phi = -\alpha H_{\rm ext}$, at which $\Delta _{\rm STT}= 0$ and $f=1/4\pi$ does not depend on $\bm{n}$, namely the probability distribution becomes isotropic over the entire Bloch sphere. For larger currents $\beta I_{\rm dc}\sin \phi < -\alpha H_{\rm ext}$, $\Delta _{\rm STT} <0$, which looks like a negative effective magnetic field that favours the magnetisation vectors in the southern Bloch hemisphere. This is a probabilistic picture of the magnetisation reversal by the STT in the presence of thermal noise. 

\begin{figure}[h]
    \centering
    \includegraphics[width=0.95\linewidth]{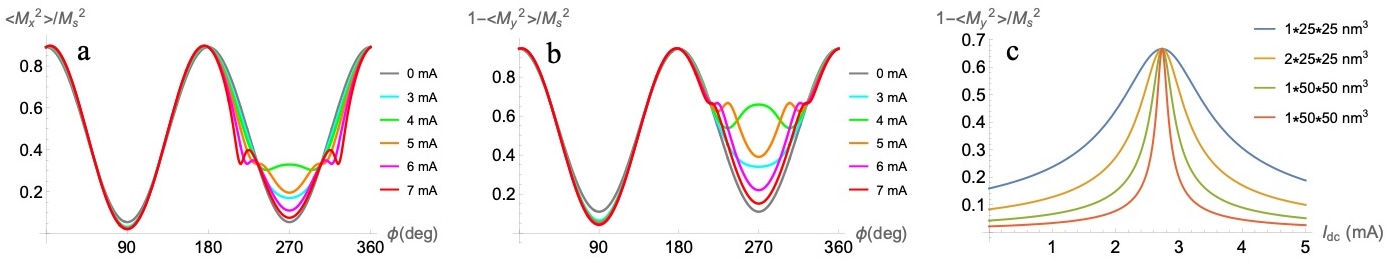}
  \caption{Field angle dependence of AMR (a) and SMR (b) for $\mu _0 H_{\rm ext}=150$~mT, $V_\text{a}=2\times25\times 25$~nm$^3$, and different current values, and current dependence of SMR (c) for $\mu _0 H_{\rm ext}=100$~mT and different values of $V_\text{a}$. The other parameters are the same as specified in Methods of the main text.}
  \label{fig:FP}
\end{figure}

Using the probability distribution, the first and second moments of the magnetisation components read
\begin{equation}
    \begin{pmatrix}
        \langle M_{x} \rangle \\
        \langle M_{y} \rangle \\
        \langle M_{z} \rangle \\
    \end{pmatrix} = M_{\rm s} \left( \coth \Delta _{\rm STT} - \frac{1}{\Delta _{\rm STT} }\right) \begin{pmatrix}
        \cos \phi \\
        \sin \phi \\
        0 \\
    \end{pmatrix},
\end{equation}
and
\begin{equation}
    \begin{pmatrix}
        \langle M_{x}^2 \rangle \\
        \langle M_{y}^2 \rangle \\
        \langle M_{z}^2 \rangle \\
    \end{pmatrix}= M_{\rm s}^2 \begin{pmatrix}
        \cos ^2 \phi \\
        \sin ^2 \phi \\
        0 \\
    \end{pmatrix} + \frac{M_{\rm s}^2}{\Delta _{\rm STT} }\left( \coth \Delta _{\rm STT} -\frac{1}{\Delta _{\rm STT}} \right) \begin{pmatrix}
        1-3\cos ^2 \phi \\
        1-3\sin ^2 \phi \\
        1 
    \end{pmatrix} , \label{eq:SMR_theory}
\end{equation}
where the angled brackets denote the thermal averaging. $\coth \Delta _{\rm STT} -1/\Delta _{\rm STT} \rightarrow \Delta _{\rm STT} /3$ as $\Delta _{\rm STT} \rightarrow 0$ confirms the isotropic magnetisation distribution at the critical current. Since AMR $\propto \langle M_{x}^2 \rangle$ and SMR $\propto M_{\rm s}^2 -\langle M_{y}^2 \rangle $ we expect different results for these mechanisms, even for an in-plane angle ($\phi $) scan, as can be seen in Figs.~\ref{fig:FP}a-b. The SMR agrees better with the experimental results. $\coth \Delta $ in Eq.~(\ref{eq:SMR_theory}) indicates an exponential dependence on temperature, and the macrospin volume $V_\text{a}$. The $I_{\rm dc}$ dependence of the SMR for different values of $V_\text{a}/T$ in Fig.~\ref{fig:FP}c compared with the rather gradual variation seen in the experiments fixes the order of magnitude of $V_\text{a}/T$.

\newpage
\section*{Supplementary Note \rev{7}: First Passage Time}

The stochastic dynamics of the magnetisation 
on the two-dimensional Bloch sphere can be can be characterized by fast precessional motion and slow drift-diffusion in energy, both of which are one-dimensional, if there is a hierarchy between the energy-conserving deterministic torques (\textit{e.g.} the magnetic field and anisotropy) and the damping, STT, and fluctuations that all change the energy~\cite{Bertotti2006x}. The magnetisation switching rate is then related to the inverse first passage time~\cite{Newhall2013x, Taniguchi2013x}. Here, we apply the same technique to calculate the dwelling time in the regime of dynamical bistability.

We assume $\bm{H}_{\rm ext}$ and the polarisation of the spin current to be parallel with $\bm{H}_{\rm ext} =H_{\rm ext}\hat{\bm{y}}$ and $I_{\rm dc}\rightarrow I_{\rm dc}\sin \phi $. Our model then reduces to a special case of Newhall and Vanden-Eijnden's work~\cite{Newhall2013x} with $\beta _y =0$, $\beta _z  \neq 1/2$, and $x \leftrightarrow y$ in their notation. While they normalised the energy and time by $\mu _0 M_{\rm s}^2 V_\text{a}$ and $1/\gamma \mu _0 M_{\rm s}$ respectively, it suits us better to normalise them by $\mu _0 M_{\rm s} H_{\rm ext}V_\text{a}$ and $1/\gamma \mu _0 H_{\rm ext} $, denoted by dimensionless variables $\epsilon$ and $\tau$. The stochastic equation after averaging of the stochastic LLG over the fast energy-conserving dynamics is
\begin{equation}
    \frac{d\epsilon }{d\tau } = -\alpha A\left( \epsilon \right) - \eta _{\rm STT} B\left( \epsilon \right) +\frac{2\alpha }{\Delta} C\left( \epsilon \right) + \sqrt{\frac{2\alpha }{\Delta } A\left( \epsilon \right)} \zeta \left( \tau \right) ,
    \label{eq:mdynamics}
\end{equation}
where $\eta _{\rm STT}=\beta I_{\rm dc}\sin \phi /H_{\rm ext}, \Delta = \mu _0 M_{\rm s} H_{\rm ext}V_{\text{a}}/k_B T $, $\zeta \left( \tau \right) $ is the normalised white noise satisfying $\langle \zeta \left( \tau \right) \zeta \left( \tau ^{\prime } \right) \rangle = \delta \left( \tau -\tau ^{\prime } \right)$, and the coefficients read
\begin{eqnarray}
    A\left( \epsilon \right) &=& \nu ^2 \frac{\overline{M_{z}^2}}{M_{\rm s}^2} -4\epsilon ^2 -4\epsilon \frac{\overline{M_{y}}}{M_{\rm s}} +1 - \frac{\overline{M_{y}^2}}{M_{\rm s}^2} , \label{eq:coeffA} \\
    B\left( \epsilon \right) &=& 2\epsilon \frac{\overline{M_{y}}}{M_{\rm s}} +1+ \frac{\overline{M_{y}^2}}{M_{\rm s}^2} ,\label{eq:coeffB} \\
    C\left( \epsilon \right) &=& \frac{\nu }{2} -3\epsilon - 2\frac{\overline{M_{y}}}{M_{\rm s}} . \label{eq:coeffC}
\end{eqnarray}
We do not have to worry about different branches~\cite{Newhall2013x} since the energy fixes the orbits unambiguously. The overbars denote averaging over a cycle of the energy-conserving fast dynamics:
\begin{eqnarray}
    \frac{\overline{M_{y}}}{M_{\rm s}} &=& \frac{1}{\tau _{\circ }\nu}\int _0^{2\pi }\frac{\sqrt{1-2\epsilon \nu \sin ^2 \phi + \nu ^2 \sin ^4 \phi }-1}{\sin ^2 \phi + \left( \sqrt{1-2\epsilon \nu \sin ^2 \phi + \nu ^2 \sin ^4 \phi }-1 \right) \cos ^2 \phi }d\phi , \label{eq:average_y} \\
    \frac{\overline{M_{y}^2}}{M_{\rm s}^2} &=& \frac{1}{\tau _{\circ }\nu ^2} \int _0^{2\pi } \frac{\left( \sqrt{1-2\epsilon \nu \sin ^2 \phi + \nu ^2 \sin ^4 \phi }-1 \right) ^2}{\sin ^2 \phi + \left( \sqrt{1-2\epsilon \nu \sin ^2 \phi + \nu ^2 \sin ^4 \phi }-1 \right) \cos ^2 \phi }\frac{d\phi }{\sin ^2 \phi }, \label{eq:average_y2}  \\
    \frac{\overline{M_{z}^2}}{M_{\rm s}^2} &=& \frac{1}{\tau _{\circ }\nu ^2} \int _0^{2\pi} \frac{\nu ^2 \sin ^4 \phi - \left( \sqrt{1-2\epsilon \nu \sin ^2 \phi + \nu ^2 \sin ^4 \phi }-1 \right) ^2}{\sin ^2 \phi + \left( \sqrt{1-2\epsilon \nu \sin ^2 \phi + \nu ^2 \sin ^4 \phi }-1 \right) \cos ^2 \phi}d\phi , \label{eq:average_z2}
\end{eqnarray}
with period
\begin{equation}
    \tau _{\circ } = \int _0^{2\pi }\frac{\sin ^2 \phi}{\sin ^2 \phi +\left( \sqrt{1-2\epsilon \nu \sin ^2 \phi + \nu ^2 \sin ^4 \phi }-1 \right) \cos ^2 \phi } d\phi . \label{eq:period}
\end{equation}

\begin{figure}[h]
    \centering
    \includegraphics[width=0.6\linewidth]{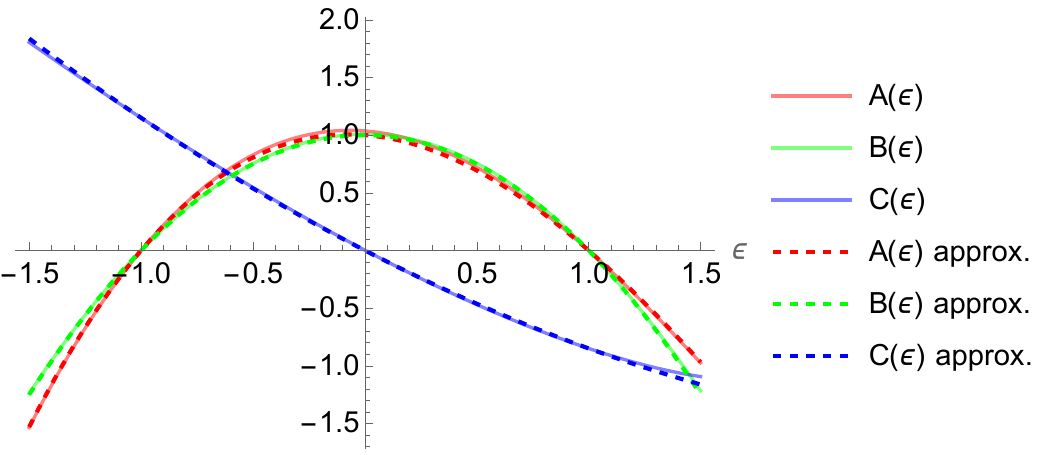}
  \caption{The coefficient functions Eqs.~(\ref{eq:coeffA})~-~(\ref{eq:coeffC}) (solid lines) and their approximations Eqs.~(\ref{eq:coeffA_approx})~-~(\ref{eq:coeffC_approx}) (dashed lines) for $\nu = M_{\rm eff}/H_{\rm ext}=0.3$.}
  \label{fig:fpt_coefficients}
\end{figure}

According to Feller~\cite{Feller1954x}, the first passage time $\tau \left( x;\epsilon _1,\epsilon _2 \right)$ from $x \in \left( \epsilon _1 ,\epsilon _2 \right)$ to one of the boundaries $x=\epsilon _1 $ or $x=\epsilon _2$ obeys
\begin{equation}
    \left\{ -\alpha A\left( x\right) -\eta _{\rm STT}B\left( x\right) + \frac{2\alpha }{\Delta } C\left( x\right) \right\} \frac{d\tau }{dx} + \frac{\alpha }{\Delta }A\left( x\right) \frac{d^2 \tau }{dx^2} = -1 ;  \label{eq:first_passage}
\end{equation}
The absorbing boundary conditions $\tau \left( \epsilon _1 \right) = \tau \left( \epsilon _2 \right) = 0$ make sense when both $\epsilon _1 ,\epsilon _2$ are accessible boundaries. However, we are interested in the passage to and from the north and south poles $x=\mp 1$, which were claimed to be entrance boundary points~\cite{Newhall2013x} with $A\left( \pm 1\right) =0$, and unreachable from any interior point $x\in \left( -1,1\right)$ in a finite time. However, we first show that $\epsilon =\pm 1$ are in fact accessible boundary points at least in our model with $0<\left| \nu \right| <1$ as follows. Let $F\left( x\right)$ be an indefinite integral defined by
\begin{equation}
    F\left( x\right) = \Delta \int ^x \left\{ -1 - \frac{\eta _{\rm STT}}{\alpha }\frac{B\left( y\right)}{A\left( y\right)} + \frac{2}{\Delta }\frac{C\left( y\right)}{A\left( y\right)} \right\} dy .
\end{equation}
A boundary point $x_0$ is accessible if the functions $e^{-F\left( x\right)}$ and $A\left( x\right) ^{-1} e^{F\left( x\right)}$ are both integrable in some neighbourhood of $x_0$~\cite{Feller1954x}. One may numerically check that $A\left( x\right) ,B\left( x\right) ,C\left( x\right) $ are all smooth functions in an open interval containing $\left[ -1,1\right]$ with $x=\pm 1$ the only zeros of $A\left( x\right)$ (Fig.~\ref{fig:fpt_coefficients}). Therefore, we just need to check the integrability around $x_0 =\pm 1$. The integrals Eqs.~(\ref{eq:average_y}) - (\ref{eq:period}) and their derivatives can be analytically evaluated for $\epsilon = \pm 1$ as $A\left( \pm 1\right) = 0 , B\left( \pm 1\right) =0 , C \left( \pm 1\right) = \mp \left( 2\mp \nu \right) /2$ and 
\begin{eqnarray}
    A\left( \epsilon \right) &=&  \left( 2\mp \nu +\frac{\nu ^2}{2\mp \nu } \right) \left( 1 \mp \epsilon \right) + o\left( 1\mp \epsilon \right) , \\
    B\left( \epsilon \right) &=& 2\left( 1\mp \epsilon \right) + o\left( 1\mp \epsilon \right) .
\end{eqnarray}
Thus one obtains
\begin{equation}
    F\left( x\right) =  \ln \left( 1\mp x \right) ^a + {\rm regular \ terms.}, \quad a =\left\{ 1 + \left( \frac{\nu }{2\mp \nu } \right) ^2 \right\} ^{-1} < 1 .
\end{equation}
Therefore, $e^{-F\left( x\right)} \sim \left( 1\mp x\right) ^{-a}$ and $A\left( x\right) ^{-1}e^{F\left(  x\right)} \sim \left( 1\mp x\right) ^{a-1}$ around $x=\pm 1$, suggesting they are integrable. 

Once the poles are accessible in a finite time, some boundary conditions other than $\tau \left( \pm 1 \right) =0$ have to be imposed to describe how the magnetisation returns back from the boundaries. While the natural choice is the reflecting lateral conditions~\cite{Feller1954x}, we avoid this consideration by taking advantage of our interest in the small $\nu$ regime. Namely, the integrals Eqs.~(\ref{eq:average_y}) - (\ref{eq:period}) can be approximately computed as
\begin{eqnarray}
    \tau _{\circ } &=& \frac{2\pi }{\sqrt{1-\epsilon \nu }} + O\left( \nu ^2 \right) , \\
    \frac{\overline{M_{y}}}{M_{\rm s}} &=& -\epsilon + \frac{1-\epsilon ^2}{4}\nu + O\left( \nu ^2 \right) , \\
    \frac{\overline{M_{y}^2}}{M_{\rm s}^2} &=& \epsilon ^2 - \frac{1-\epsilon ^2}{2}\epsilon \nu + O\left( \nu ^2 \right) , \\
    \frac{\overline{M_{z}^2}}{M_{\rm s}^2} &=&\frac{1-\epsilon ^2}{2} \left( 1 + \frac{\epsilon \nu }{2} \right) +O\left( \nu ^2 \right) .
\end{eqnarray}
Therefore the coefficient functions in Eqs.~(\ref{eq:coeffA})~-~(\ref{eq:coeffC}) read
\begin{eqnarray}
A\left( \epsilon \right) & \approx & \left( 1-\epsilon ^2 \right) \left( 1-\frac{\epsilon \nu }{2} \right)  , \label{eq:coeffA_approx} \\
B\left( \epsilon \right) &\approx & 1-\epsilon ^2 , \\
C\left( \epsilon \right) &\approx & -\epsilon + \frac{\nu \epsilon ^2}{2}  .\label{eq:coeffC_approx}
\end{eqnarray}
As can be seen in Fig.~\ref{fig:fpt_coefficients}, the difference from the exact result is barely noticeable and we adopt the approximated equation as our physical model. At this first order in $\nu $, it turns out that $e^{-F\left(  x\right)}$, constructed from the approximate coefficient functions, is not integrable, which implies the boundaries $x =\pm 1$ are inaccessible. One can then expect the following:
\begin{enumerate}
    \item If one of $\epsilon _1 \rightarrow -1$ or $\epsilon _2 \rightarrow 1$ is taken, it is expected that the probability of the system reaching that boundary goes to zero and $\tau \left( x;\epsilon _1 ,\epsilon _2 \right)$ should give the first passage time from $x$ to $\epsilon _2 < 1$ or $\epsilon _1 >-1 $ accordingly. 
    \item If both $\epsilon _1 \rightarrow -1 , \epsilon _2 \rightarrow 1$ are taken, $\tau \left( x\right) \rightarrow \infty $ is expected for every $x$.
\end{enumerate}
\begin{figure}[t]
    \centering
    \includegraphics[height=6cm]{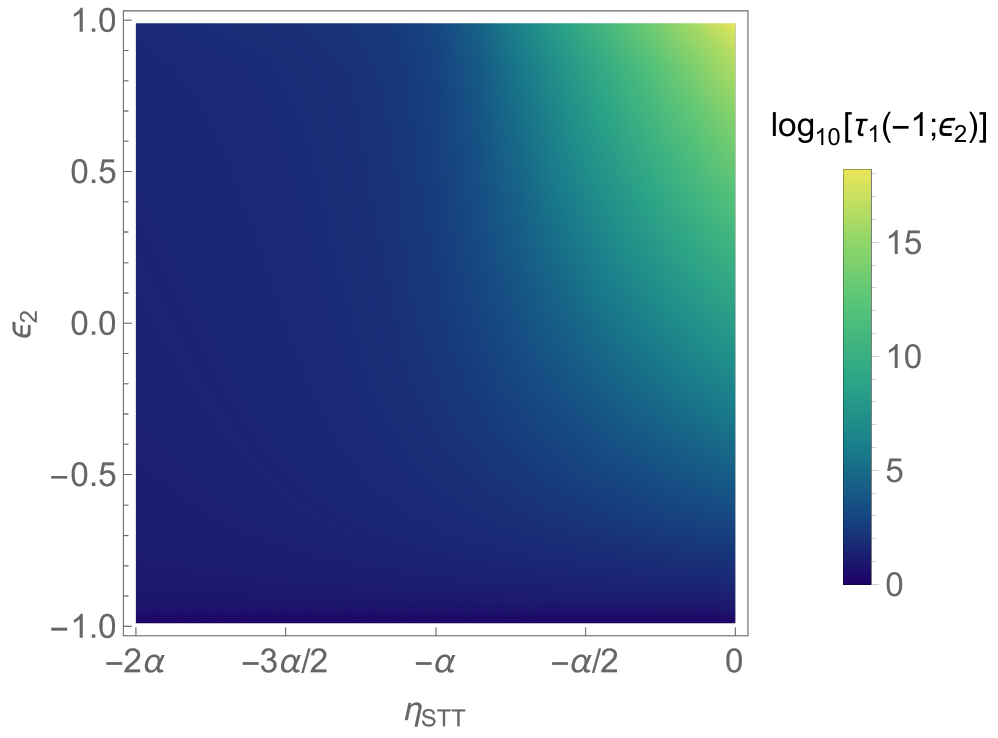}
    \includegraphics[height=6cm]{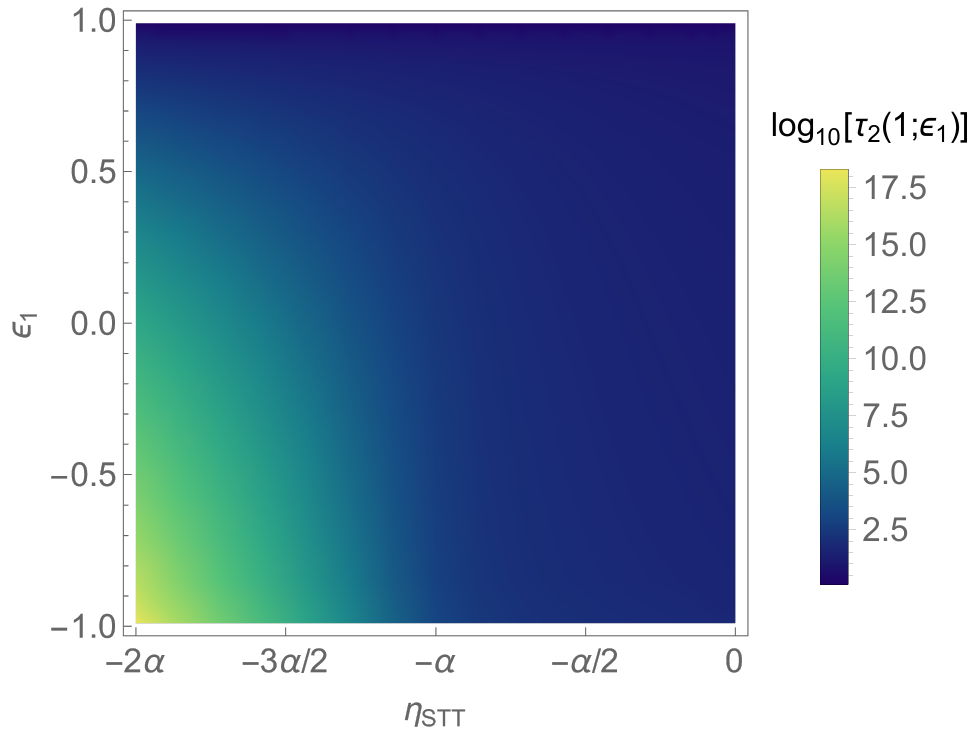}
  \caption{First passage times for $\alpha =0.09, \nu = 0.2$ from south to north $\lim _{x\rightarrow -1}\tau _1 \left( x;\epsilon _2 \right) $ (left panel) and north to south $\lim _{x\rightarrow +1}\left( x;\epsilon _1 \right)$ (right panel) as functions of $\eta _{\rm STT}$ and the end point \textit{i.e.}, $\epsilon _2$ for $\tau _1$ and $\epsilon _1$ for $\tau _2$.}
  \label{fig:fpt_maps}
\end{figure}
Therefore, in order to estimate the switching time from the north pole to the south and \textit{vice versa}, we calculate $\tau _1 \left( x; \epsilon _2 \right) \equiv \lim _{\epsilon _1 \rightarrow -1}\tau \left( x;\epsilon _1 ,\epsilon _2 \right)$ and $\tau _2 \left( x;\epsilon _1 \right) \equiv \lim _{\epsilon _2 \rightarrow 1}\tau \left( x;\epsilon _1 ,\epsilon _2 \right) $. We then take the limit $x\rightarrow -1$ for an arbitrarily $\epsilon _2 \lesssim 1$ and $x\rightarrow 1 $ for an arbitrarily $\epsilon _1 \gtrsim -1$. For this programme, we first rewrite Eq.~(\ref{eq:first_passage}) as
\begin{equation}
    \frac{d}{dx}\left\{ e^{F\left( x\right)} \frac{d\tau }{dx} \right\} = -\frac{\Delta e^{F\left( x\right)}}{\alpha A\left( x\right)}. \label{eq:first_passage_2}
\end{equation}
The integration constant of $F$ cancels in Eq.~(\ref{eq:first_passage_2}), and the right-hand-side is integrable. When solving Eq.~(\ref{eq:first_passage_2}), the integration constants do have to be specified such that the boundary conditions $\tau \left( \epsilon _1 \right) = \tau \left( \epsilon _2 \right) = 0$ are satisfied. Since we eventually take $\epsilon _1 \rightarrow -1$ for $\tau _1$ and $\epsilon _2 \rightarrow 1 $ for $\tau _2$ respectively, it is more convenient to make different choices between the two
\begin{eqnarray}
    \tau _1 \left( x;\epsilon _2 \right) &=&  \int _x^{\epsilon _2} e^{-F\left( y\right)} \left\{ G_1 \left( y\right) - C_1 \right\} dy , \quad G_1 \left( y\right) \ = \ -\frac{\Delta }{\alpha }\int _y^{\epsilon _2}\frac{e^{F\left( z\right)}}{A\left( z\right)}dz , \\
    \tau _2 \left( x;\epsilon _1 \right) &=& \int _{\epsilon _1}^x e^{-F\left( y\right)} \left\{ G_2 \left( y\right) - C_2 \right\} dy , \quad G_2 \left( y\right) \ = \ -\frac{\Delta }{\alpha } \int _{\epsilon _1}^y \frac{e^{F\left( z\right)}}{A\left( z\right)}dz ,
\end{eqnarray}
which automatically satisfy $\tau _1 \left( \epsilon _2 ;\epsilon _2 \right) = \tau _2 \left( \epsilon _1 ;\epsilon _1 \right) =0 $ and the dependence on $\epsilon _1$ for $\tau _1$ or $
\epsilon _2$ for $\tau _2$ is contained entirely in $C_1$ or $C_2$ respectively. Note that $G_1\left( y\right) , G_2 \left( y \right)$ are bounded in $\left[ -1 ,1\right]$. The remaining integration constants $C_1 ,C_2$ are chosen so as to satisfy $\tau _1 \left( \epsilon _1 ;\epsilon _2 \right) =\tau _2 \left( \epsilon _2 ;\epsilon _1 \right) = 0$, yielding
\begin{eqnarray}
C_1 \left( \epsilon _1 \right) &=& \int _{\epsilon _1}^{\epsilon _2} e^{-F\left( y\right)} G_1 \left( y\right) dy \bigg/\int _{\epsilon _1}^{\epsilon _2}e^{-F\left( y\right)}dy , \\
C_2 \left( \epsilon _2 \right) &=& \int _{\epsilon _1}^{\epsilon _2}e^{-F\left( y\right)}G_2 \left( y\right) dy \bigg/ \int _{\epsilon _1}^{\epsilon _2} e^{-F\left( y\right)} dy.
\end{eqnarray}
When the limit $\epsilon _1 \rightarrow -1$ for $C_1$ or $\epsilon _2 \rightarrow 1$ for $C_2$ is taken, the integrals both in the numerator and denominator will be dominated by the region $y\approx -1$ or $y\approx 1$ respectively, as $e^{-F\left( y\right)} \sim \left( 1\pm y\right) ^{-1}$. Thus, one can guess that $\lim _{\epsilon _1 \rightarrow -1} C_1 \left( \epsilon _1 \right) = G_1 \left( -1\right)$ and $\lim _{\epsilon _2 \rightarrow 1}C_2\left( \epsilon _2 \right) = G_2 \left( 1\right)$. We accept them without a proof and obtain
\begin{eqnarray}
    \tau _1 \left( x;\epsilon _2 \right) &=& \frac{\Delta }{\alpha } \int _x^{\epsilon _2} e^{-F\left( y\right)}\left\{ \int _{-1}^y \frac{e^{F\left( z\right)}}{A\left( z\right)}dz \right\} dy \ \approx \ \frac{\Delta }{\alpha }\int _x^{\epsilon _2}\int _{-1}^y \frac{1}{1-y^2}\left( \frac{2-\nu z}{2-\nu y} \right) ^{\frac{2\eta _{\rm STT} \Delta }{\alpha \nu }} \left( 1-\frac{\nu z}{2} \right) ^{-1} e^{\Delta \left( y-z\right)}dz dy ,\label{eq:t1}  \\
    \tau _2 \left( x;\epsilon _1 \right) &=& \frac{\Delta }{\alpha }\int _{\epsilon _1}^x e^{-F\left( y\right)} \left\{ \int _y^1 \frac{e^{F\left( z\right)}}{A\left( z\right)}dz \right\} dy \ \approx \ \frac{\Delta }{\alpha }\int _{\epsilon _1}^x \int _y^1 \frac{1}{1-y^2}\left( \frac{2-\nu z}{2-\nu y} \right) ^{\frac{2\eta _{\rm STT} \Delta }{\alpha \nu }} \left( 1-\frac{\nu z}{2} \right) ^{-1} e^{\Delta \left( y-z\right)} dz  dy. \label{eq:t2}
\end{eqnarray}
We remark that although the argument about the limits of $C_1 , C_2$ does not hold when the boundaries are accessible, the expressions Eqs.~(\ref{eq:t1}) and (\ref{eq:t2}) before substituting the approximations in fact agree with what they should be under the reflecting lateral conditions. 

The integrals are evaluated and plotted in Fig.~\ref{fig:fpt_maps}. The anisotropy $\nu =0.2$ models $\mu _0 M_{\rm eff}=30$~mT and $\mu _0 H_{\rm ext}=150$~mT. To compare the result with the dwelling time observed in Fig.~4e in the main text, we note that the first passage times are $\tau _1 \sim \tau _2 \sim 10^3$ for $\eta _{\rm STT}=-\alpha$. For $\mu _0 H_{\rm ext}=150$~mT, $1/\gamma \mu _0 H_{\rm ext}=0.24$~ns so that the theory predicts a dwelling time of order microseconds in a qualitative agreement with the stochastic simulation. 
\newpage
\section*{Supplementary Note \rev{8}: Joule Heating}

We include the effect of Joule heating by a temperature-dependent anisotropy and magnetisation. The magnetisation of W\rev{\textbar}CoFeB(1.3~nm)\rev{\textbar}MgO in Fig.~2b of Ref.~\citeonline{Lee_AIPAdv_7_065107_2017} is well represented by Kuz{\textquoteright}min's equation~\cite{Kuz_min_PhysRevLett_94_107204_2005x}:
\begin{equation}
    m(T) = \left(1 - s(T/T_{\mathrm{C}})^{3/2} - (1-s)(T/T_{\mathrm{C}})^p \right)^{1/3}
\end{equation}
with $p=2.5$, $s=2.5$ and $T_{\mathrm{C}} = 833$~K. The interface anisotropy scales as $K(T)/K(0) = (M(T)/M(0))^2$ according to the Callen-Callen theory~\cite{Callen_JPhysChemSolids_27_1271_1966x}. The magnetisation and anisotropy at a given temperature are:
\begin{equation}
    M_{\rm s}(T) = M_{\rm s}\frac{m(T/T_{\mathrm{C}})}{m(300K)}\quad K(T) = K\frac{m(T/T_{\mathrm{C}})^2}{m(300K)^2}.
\end{equation}
Using the electrical resistance as a temperature sensor, we estimate $\Delta T_{\mathrm{max}} = 100$~K for the maximum current $j_{\mathrm{max}}=8$~mA and interpolate for the temperature increase as
\begin{equation}
    T = T_{\mathrm{ambient}} + \Delta T_{\mathrm{max}} \left(\frac{j}{j_{\mathrm{max}}}\right)^2.
\end{equation}

\newpage
\section*{Supplementary Note \rev{9}: Continuous restricted Boltzmann machines using nearly isotropic magnets}

Here we show probabilistic computing applications using the stochasticity of an array of nearly-isotropic magnets as a computational resource. Electrically controlled zero damping states achieved in our study represent a platform to construct probabilistic bits (p-bits) due to its stochastic transitions across the entire Bloch sphere. Our p-bits can output a continuous variable between the two states, i.e. $x \in \left[-1, 1\right]$. Such behaviour stands in stark contrast to that of binary p-bits ($x \in \{-1, 1\}$), which are based on stochastic magnetic tunnel junctions (s-MTJs) and can be correlated to minimise an Ising Hamiltonian~\cite{borders2019smtjsx,suttonpcoprocessing2020x} for binary tasks such as machine learning~\cite{Chowdhury_IEEE2023x}, combinatorial optimisation~\cite{borders2019smtjsx,Chowdhury_IEEE2023x,kampfe2025tspx} and quantum emulation~\cite{Chowdhury_IEEE2023x,chowdhury2020qmx}. P-bits can also be used for the Boltzmann machine (BM), where one can make use of the BM update rules \cite{bunaiyan2024heis} to train the synaptic connections between p-bits to sample from desired distributions.

\begin{figure}[h!]
\centering
\includegraphics[width=\linewidth]{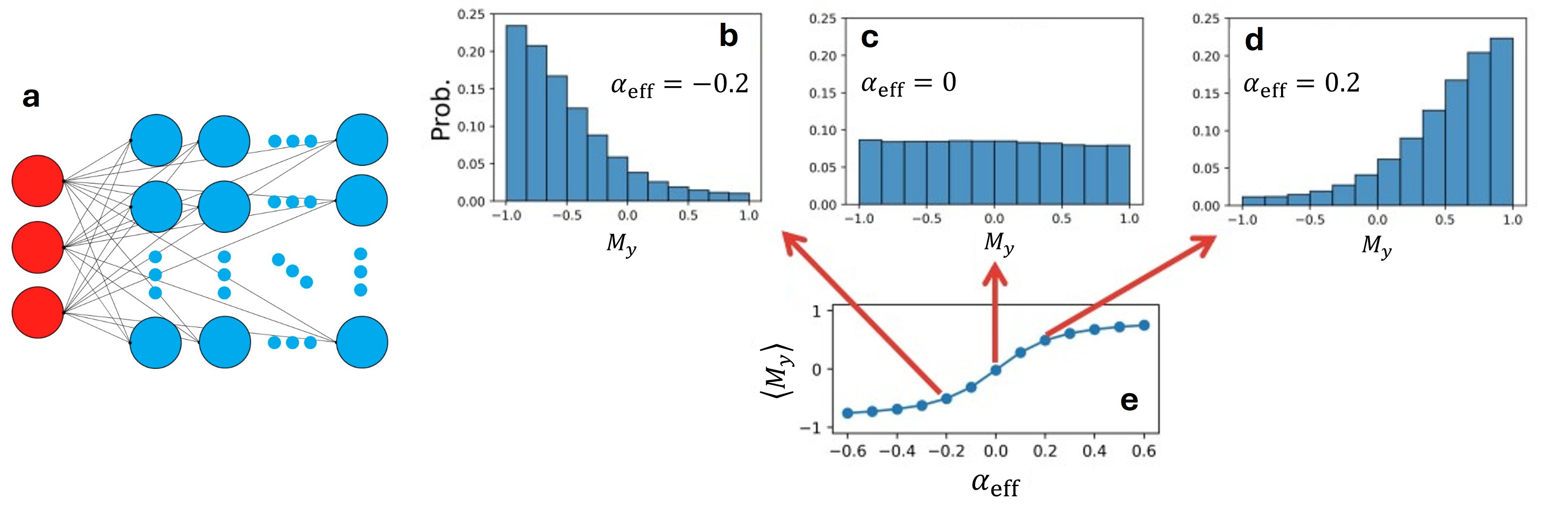}
\caption{\textbf{a,} Schematic of a RBM. Hidden units (red) are synaptically connected to visible units (blue). \textbf{b-d,} Histograms showing $M_y$ values for different effective damping parameters. \textbf{e,} Time-averaged $M_y$ for varied $\alpha_{\text{eff}}$. Simulations were performed using the Euler method with the following parameters: time-step 0.1 ps (made under the assumption of a 100mT external field), total simulation time 1 $\mu$s, $V_\text{a}=1\times4\times4$~nm$^3$, $\alpha=0.1$ and $M_{\rm eff}/H_{\rm ext}=0$.}
\label{fig:tanh}
\end{figure}

Restricted Boltzmann machines (RBMs) offer an effective architecture to generate binary data~\cite{salakhutdinov2007rbmx,bereux2025rbmx}. RBMs comprise two groups of binary stochastic variables $\{0,1\}$: \textit{visible} p-bits ($\textbf{v}$) and \textit{hidden} p-bits ($\textbf{h}$) that are connected by weights ($\textbf{W}$), as schematically shown in Fig.~\ref{fig:tanh}a. Both layers are also influenced by visible and hidden bias vectors, $\boldsymbol{\chi}$ and $\boldsymbol{\rho}$ respectively. The Boltzmann distribution for this network can therefore be expressed as~\cite{bereux2025rbmx}:

\begin{equation}
    p(\textbf{v,~h})=\frac{1}{Z} \text{e}^{-H(\textbf{v,~h})}
\end{equation}

\noindent where Z is the partition function and the Hamiltonian, $H(\textbf{v,~h})$, is defined as:

\begin{equation}
    H(\textbf{v,~h})=-\sum_{ij}v_{i}W_{ij}h_{j}-\sum_{i}\chi_{i}v_{i}-\sum_{j}\rho_{j}h_{j}
\end{equation}

\noindent The binary magnetisation state ($m$) of an s-MTJ at time $t$ is represented by \cite{camsari2017p_bitsx}:

\begin{equation} \label{eqn:binary_state}
    m(t)=\text{sign[rand(-1,1)+tanh[}I(t)]]
\end{equation}

\noindent where $\text{rand(-1,1)}$ is a uniform random number between -1 and 1, and $I(t)$ is the input current which can bias the s-MTJ state into either configuration via spin transfer torque. The training procedure for a RBM uses Gibbs Sampling as follows~\cite{bunaiyan2024heisx}:

\begin{enumerate}
    \item Randomly initialise the weight matrix according to $W \sim \mathcal{N}(0,0.1)$ and set the bias vectors, $\boldsymbol{\chi}$ and $\boldsymbol{\rho}$, to zero.
    \item Input a data sample to the visible layer by clamping the visible units, $\mathbf{v}^{0}$, to a data point (e.g. an image from the Fashion MNIST dataset). 
    \item Sample the hidden units using the conditional probability: $\mathbf{h}^{(0)}\sim P(\mathbf{h}^{(0)}=1|\mathbf{v}^{(0)})$.
    \item Sample the visible units using the conditional probability: $\mathbf{v}^{(1)}\sim P(\mathbf{v}^{(1)}=1|\mathbf{h}^{(0)})$.
    \item Sample the hidden units using the conditional probability: $\mathbf{h}^{(1)}\sim P(\mathbf{h}^{(1)}=1|\mathbf{v}^{(1)})$.
    \item Calculate the weight and bias updates using contrastive divergence: $\Delta\mathbf{W}=\eta[\mathbf{v}^{(0)}\mathbf{h}^{(0)\text{T}} - \mathbf{v}^{(1)}\mathbf{h}^{(1)\text{T}}]$, $\Delta\boldsymbol{\chi}=\eta[\mathbf{v}^{(0)} - \mathbf{v}^{(1)}]$ and $\Delta\boldsymbol{\rho}=\eta[\mathbf{h}^{(0)} - \mathbf{h}^{(1)}]$, where $\eta$ is the learning rate hyperparameter.
    \item Apply the updates to the weight matrix and bias vectors.
    \item Repeat steps 2-7 for each data point in the training data batch and monitor the model's performance using the metrics shown in Table~\ref{tab:metrics}.
\end{enumerate}

\noindent Inference is subsequently performed using the optimised model parameters (with initially randomised visible and hidden units) to repeat steps 3 and 4, $N$ times, to generate new data points. It is possible to amend this binary model to output continuous data by taking an average value of each visible p-bit, yielding values on the interval [0,1]. This augmentation is termed a rate-coded RBM (RBMrate)~\cite{chen2002crbmx,chen2003crbmx}, and can be physically modelled using established s-MTJ p-bits~\cite{Chowdhury_IEEE2023x}. However, RBMrate models are limited in their implementation as repetitive sampling is unfeasible in hardware for large p-circuits~\cite{chen2002crbmx,chen2003crbmx}. As well as this, RBMrate is intrinsically prone to mode collapse in their generated samples, as averaging across many binary samples for each continuous output, leads to a loss in data diversity~\cite{chen2002crbmx}. 

In order to process continuous data effectively, one must replace the binary operation of the model architecture with a continuous functionality, such that all p-bits exist on the interval [0,1] at every timestep. This architecture is known as a continuous restricted Boltzmann machine (cRBM) and has been extensively researched using software approaches~\cite{chen2002crbmx,chen2003crbmx,PARRA_NeuralNetwork1995x,harrison2018continuousrbx}. A cRBM fundamentally differs from an RBM/RBMrate by replacing the binary stochastic unit [described by Eq.(\ref{eqn:binary_state})] with a continuous stochastic unit, where the state of the $j^{\text{th}}$ unit, $s_{j}$, is defined by \cite{chen2002crbmx}:

\begin{equation}
    s_{j}=\varphi\Big[\sum_{i} W_{ij}s_{i}+\sigma \times N_{j}(0,1)\Big]
\end{equation}

\noindent $\varphi$ is a sigmoid function and $\sigma \times N_{j}(0,1)$ is a Gaussian sample with standard deviation $\sigma$. To physically realise this continuous stochastic unit using the MgO\rev{\textbar}CoFeB\rev{\textbar}W stack, we operate in the isotropic limit, i.e. $M_{\text{eff}}/H_{\text{ext}}\rightarrow0$, allowing the system to evolve as a random walk (following Eq.~\ref{eq:mdynamics}) about the axis parallel to the external field (y-axis) at finite temperatures. We can bias the system by tuning the effective damping: $\alpha_{\text{eff}}=\alpha-\beta I_{\rm dc}\sin \phi /H_{\rm ext}$ where $\beta I_{\rm dc}\sin \phi /H_{\rm ext}$ is the field normalised strength of the spin-current induced anti-damping torque. Figures~\ref{fig:tanh}b-d represent the probability distribution of <$M_y$> for different $\alpha_{\text{eff}}$, with the $\alpha_{\text{eff}}$ dependence of its time-average over 1 $\mu$s shown in Fig.~\ref{fig:tanh}e. 

\begin{figure}[t!]
\centering
\includegraphics[width=\linewidth]{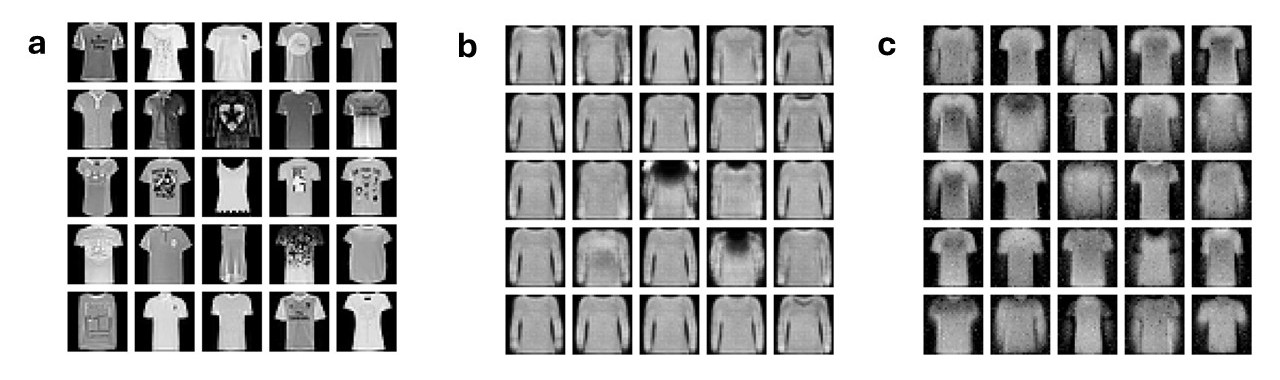}
\caption{\textbf{a,} Sample of the Fashion MNIST training data (t-shirt class). \textbf{b \& c,} Samples of generated data using RBMrate and cRBM architectures respectively. Both models comprised 10000 binary hidden p-bits and 784 binary/continuous visible units. Idealised models were trained for 5000 epochs to optimise respective model parameters. At inference, the same sLLG parameters were used for the cRBM as the data in Fig.~\ref{fig:tanh}, except for $V_\text{a}=1\times64\times64$~nm$^3$, to reduce thermal noise. Note that these figures in Fig.~S9 are identical to Fig.~5 in the main manuscript. }
\label{fig:fmist}
\end{figure}

We show the applicability of simulated isotropic MgO\rev{\textbar}CoFeB\rev{\textbar}W stacks as visible units in a cRBM for generating continuous data on the Fashion-MNIST benchmark task at inference (Fig.~\ref{fig:fmist}a), where $\alpha_{\text{eff}}$ serves as the input parameter, after training on an idealised software cRBM architecture. This training process follows the same procedure as the RBM (steps 1-8), except the conditional distributions for the visible units instead take the form $P(\mathbf{v}^{(t_{0})}|\mathbf{h}^{(t_{0})})$, for an arbitrary timestep $t_{0}$. We compare the results of RBMrate and cRBM architectures with identical parameters. Visually, one can see a greater variety in the generated "t-shirt" samples using the cRBM (Fig.~\ref{fig:fmist}c) compared to the RBMrate (Fig.~\ref{fig:fmist}b). Quantitatively, we use various metrics to measure the quality and diversity of generated samples from the cRBM and RBMrate: the Fr{\'e}chet Distance (FD) \cite{doan2020imagegenerationminimizingfrechetx}, the Multi-Scale Structure Similarity Index (MS-SSIM) \cite{wang2003msssimx}, the Jensen Shannon Divergence (JSD)~\cite{lin1991divergencex} and the Number of Statistically Different Bins (NDB)~\cite{richardson2018gansx}. The FD is a measure of similarity between two distributions that assumes both the real and generated sample sets can be accurately modelled as multivariate Gaussian distributions~\cite{Thomas2025x,doan2020imagegenerationminimizingfrechetx}. For real images ($\mathbb{P}_{r}$) and generated images ($\mathbb{P}_{g}$) with means $\mu_{r}$ and $\mu_{g}$ and covariances $\Sigma_{r}$ and $\Sigma_{g}$ respectively, the FD is defined as~\cite{doan2020imagegenerationminimizingfrechetx}:

\begin{equation}
    \text{FD}(\mathbb{P}_{r},\mathbb{P}_{g})=|\mu_{r}-\mu_{g}|^{2}+\text{Tr}\Big(\Sigma_{r}+\Sigma_{g}-2\sqrt{\Sigma_{r}\Sigma_{g}}\space\Big)
\end{equation}

\begin{table}[b!]
\centering
\begin{tabular}{|c|c|c|}
\hline
& RBMrate & cRBM \\
\hline
FD (\textit{inter}) $\downarrow$ & 0.05 & \textbf{0.02} \\
MS-SSIM (\textit{inter}) $\uparrow$ & 0.18 & \textbf{0.22} \\
MS-SSIM (\textit{intra}) $\downarrow$ & 0.54 & \textbf{0.26} \\
JSD (\textit{inter}) $\downarrow$ & 0.99 & \textbf{0.90} \\
JSD (\textit{intra}) $\uparrow$ & 0.33 & \textbf{0.67} \\
NDB/\textit{k} (\textit{inter}) $\downarrow$ & 0.97 & \textbf{0.87} \\
\hline
\end{tabular}
\caption{\textbf{Performances of RBMrate and cRBM on Fashion-MNIST t-shirt generation.} 1000 samples were used for every sample set when calculating each metric. Bold values correspond to the best scores and the arrows show whether a higher ($\uparrow$) or lower ($\downarrow$) value is optimal for each metric. $k=30$ was used when calculating NDB/\textit{k}.}
\label{tab:metrics}
\end{table}

\noindent A lower FD score suggests that the generated data more closely resembles the true data distribution \cite{Thomas2025x}. The SSIM accesses the difference in luminance, contrast and structure between two sets of image samples \cite{wang2004ssimx}, $\textbf{x}$ and $\textbf{y}$. These three comparison components are defined as~\cite{dosselmann2011ssimx}:

\begin{equation}
    l(\textbf{x, y}) = \frac{2 \mu_x \mu_y + C_1}{\mu_x^2 + \mu_y^2 + C_1}
\end{equation}

\begin{equation}
    c(\textbf{x, y}) = \frac{2 \sigma_x \sigma_y + C_2}{\sigma_x^2 + \sigma_y^2 + C_2}
\end{equation}

\begin{equation}
    s(\textbf{x, y}) = \frac{\sigma_{xy} + C_3}{\sigma_x \sigma_y + C_3}
\end{equation}

\noindent where $\mu_{x}$ and $\mu_{y}$ are the mean pixel values of $\textbf{x}$ and $\textbf{y}$, while $\sigma_{x}$ and $\sigma_{y}$ are the corresponding standard deviations and $\sigma_{xy}$ is the covariance. Also, $C_{1,2,3}$ are small stability constants to avoid division by zero. The SSIM is therefore defined as a product of these three components:

\begin{equation}
    \text{SSIM}(\textbf{x, y}) = l(\textbf{x, y}) \cdot c(\textbf{x, y}) \cdot s(\textbf{x, y})
\end{equation}

\noindent The MS-SSIM calculates the SSIM at multiple resolutions to capture fine and coarse structures and is defined as~\cite{wang2004ssimx,wang2003msssimx,dosselmann2011ssimx}:

\begin{equation}
    \text{MS-SSIM}(\textbf{x},\textbf{y})=[l_{M}(\textbf{x},\textbf{y})]^{\alpha_{M}} \prod^{M}_{j=1}[c_{j}(\textbf{x},\textbf{y})]^{\beta_{j}}[s_{j}(\textbf{x},\textbf{y})]^{\gamma_{j}}
\end{equation}

\noindent where the SSIM is evaluated at $M$ different scales and $\alpha,\beta,\gamma$ are chosen weight exponents. A perfect MS-SSIM score of 1, indicates that the two sample sets are identical, while a score of 0 suggests that there is no correlation between the sample sets. The JSD metric is extracted using feature extraction from principal component analysis (PCA) to broadcast both sets of samples to a 2D feature space. We then partition the feature space into discrete bins, before calculating the JSD using the following expression~\cite{lin1991divergencex}:     

\begin{equation}
    \text{JSD}(\textbf{x} \parallel \textbf{y}) = \frac{1}{2} D_{\text{KL}}(\textbf{x} \parallel \textbf{m}) + \frac{1}{2} D_{\text{KL}}(\textbf{y} \parallel \textbf{m})
\end{equation}

\noindent where $\textbf{m} = \frac{1}{2}(\textbf{x} + \textbf{y})
$ and the Kullback Leibler divergence is defined as: $D_{\text{KL}}(\textbf{p} \parallel \textbf{q}) = \sum_i p_i \log \left( \frac{p_i}{q_i} \right)
$. This provides a similarity measure between two distributions on the interval [0,1], where a lower score indicates that both distributions are closely related~\cite{Thomas2025x}. The NDB is a diversity and coverage metric for the generated data which quantifies how well the generated data matches the mode distribution of the real data~\cite{Thomas2025x}. We initially utilise \textit{k}-means clustering on the real data to identify \textit{k} clusters in the feature space. We then inspect how the generated data is distributed across these \textit{k} clusters, where we proceed to perform a statistical "z-test" to identify the number of statistically different clusters (bins). Practically, the NDB evaluates the diversity of the generated data by returning the number of image types that our model cannot generate. We report a normalised NDB (NDB/\textit{k}), where a lower score implies our model is capable of generating more diverse data~\cite{richardson2018gansx}.

It is clear from Table \ref{tab:metrics} that the cRBM is capable of generating samples which more closely resemble the real data and has a better mode coverage. We use MS-SSIM and JSD in two contexts: \textit{inter} and \textit{intra} where we compare a generated sample set with the real data and two generated sample sets, respectively. We only calculate the FD and NDB/\textit{k} for the \textit{inter} case, as they are not meaningful metrics for comparing data from the same origin. The cRBM's lower FD score suggests that our generated samples are more structurally similar to the real data and more diverse, compared to the RBMrate's generated data. The MS-SSIM results support this picture, as the cRBM's higher \textit{inter} score implies a better perceptual agreement with the real data and the lower \textit{intra} score indicates a greater diversity. This is reflected in the JSD scores, where the lower \textit{inter} score shows a greater similarity between the real data and the generated data while the higher \textit{intra} score suggests the cRBM has a better mode coverage. Finally, the cRBM's lower NDB/\textit{k} score shows that the cRBM is able to produce samples that match the real distribution across more modes.

In summary, the cRBM architecture using the isotropic magnets consistently performs better for every metric investigated and is capable of generating more realistic and diverse image data than the RBMrate. The explanation for this stems from two key considerations. Firstly, the cRBM's visible units are continuous, which enable the model to process data using floating point numbers at a higher fidelity than the RBMrate's binary units \cite{harrison2018continuousrbx}. Secondly, generating continuous outputs from an RBM necessitates averaging each visible p-bit over a time interval. This leads to a loss in the generated data diversity as the nuances in each binary image are integrated over to output an "averaged" image for every generated sample. The cRBM circumvents this issue as each visible unit exists in a continuous state at every time-step. These findings present a path to physically realise a cRBM using a magnetic system, which cannot be implemented with current binary s-MTJ technology.

\newpage
\section*{Further Supplementary Figures}
\begin{figure}[h!]
    \centering
    \includegraphics[width=1.0\linewidth]{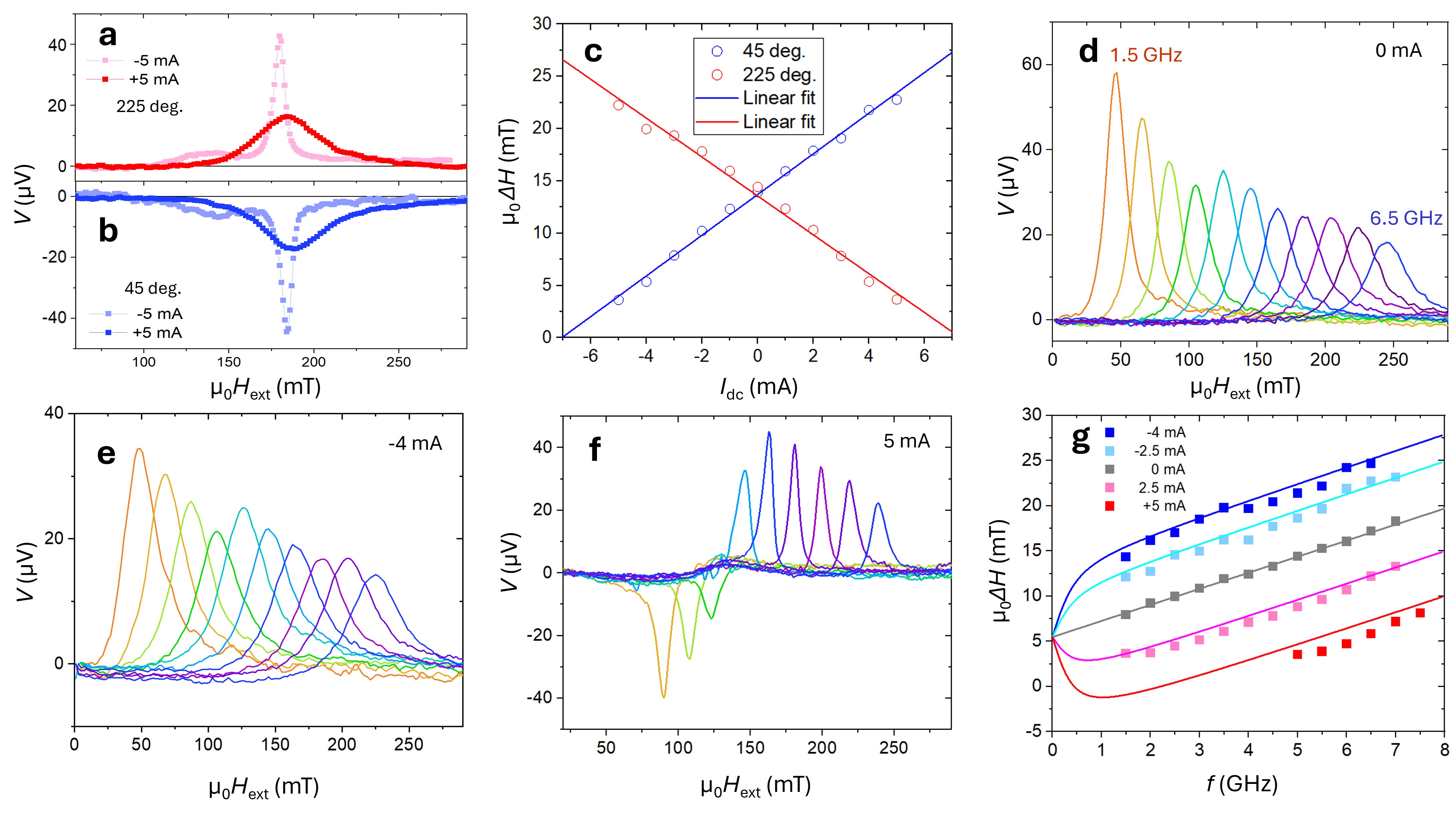}
  \caption{FMR results from another device in a different chip: \textbf{a-b}, Field-swept FMR voltages measured for (\textbf{a}) $\phi=225$ deg. and (\textbf{b}) $\phi=45$ deg.  \textbf{c}, Linewidth extracted from 5 GHz FMR measurements for $\phi=45$ deg. (blue) and  $\phi=225$ deg.(red), for different $I_\text{c}$ values. Best fit linear lines for each measurement set are also shown.  
 \textbf{d-f}, Field-swept FMR voltages measured for various frequencies and three different dc current conditions (\textbf{d} 0 mA, \textbf{e} -4 mA and \textbf{f} 5 mA). Magnetic fields were applied along $\phi=225$ deg. \textbf{g}, Linewidth as a function of frequency for different $I_\text{dc}$. The grey line is generated by a best linear fit parameter and the rest are calculated by using Eq.(3) in the main manuscript. Extracted physical parameters are $\theta _{\rm SH}$ = 0.21, $\beta=2.5\times10^6$ m$^{-1}$, $\mu_0  M_{\rm eff} =41.6$ mT, $\alpha =0.045$ and $\mu_0\Delta H_0 = 5.5$ mT.}
  \label{fig:SampleC}
\end{figure}

\begin{figure}[h!]
    \centering
    \includegraphics[width=1.0\linewidth]{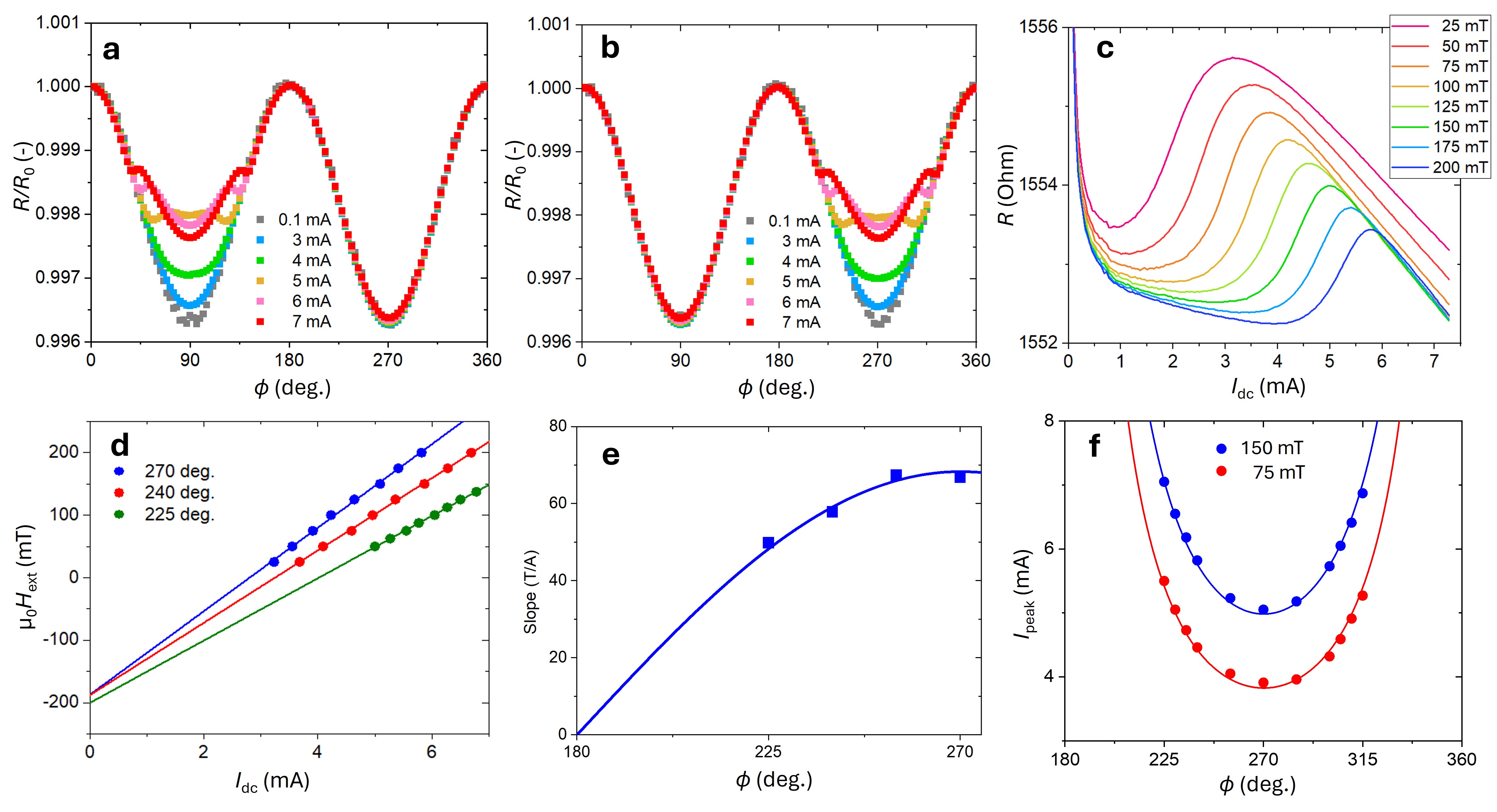}
  \caption{MR results from another device in a different chip: \textbf{a-b}, Angular dependence of magnetoresistance measured with different current biases. We applied 150~mT for all the measurements. \textbf{c}, Resistance vs dc current measurements at different applied magnetic fields 25-200~mT along $\phi=270$~deg. \textbf{d}, Peak current position measured for different magnetic fields and $\phi$, together with best-fit linear lines for each $\phi$. \textbf{e}, The slope extracted for different $\phi$ fit using a $\sin\phi$ function. \textbf{f}, Peak current position ($I_\text{peak}$) as a function of $\phi$ measured at fixed field values (75 and 150~mT) with curves calculated by Eq. (1) in the main manuscript. The best fit curves lead to $\theta_\text{SH}/\alpha = 4.6$; with $\alpha =0.045$ extracted above, $\theta _{\rm SH}$ = 0.21 is calculated and agrees very well with that quantified by FMR experiments; $\mu_0 \Delta H_0^{\prime} = 8.0$~mT is compared well with $\mu_0\Delta H_0 = 5.5$ mT.}
  \label{fig:SampleC_mr}
\end{figure}

\begin{figure}[th!]
    \centering
    \includegraphics[width=0.75\linewidth]{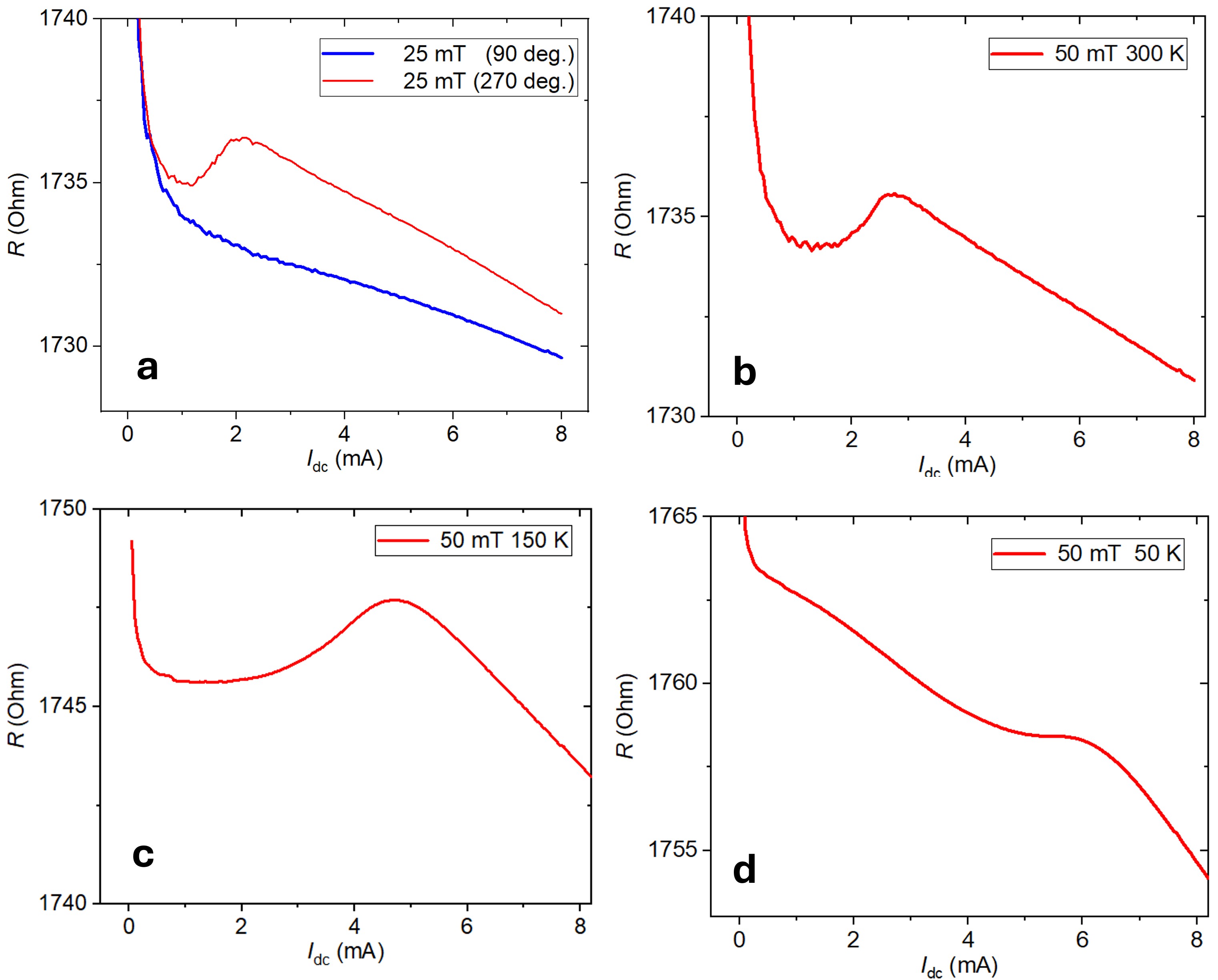}
  \caption{\textbf{a}, Magnetoresistance measurements as a function of $I_\text{dc}$, at room temperature, applying 25 mT along $\phi$ = 270 and 90 deg. respectively.  \textbf{b}-\textbf{d}, magnetoresistance measurements applying 50~mT along $\phi$ = 270 deg. for different temperatures 300, 150 and 50 K respectively.}
  \label{fig:lowT}
\end{figure}

\begin{figure}[h!]
    \centering
    \includegraphics[width=0.75\linewidth]{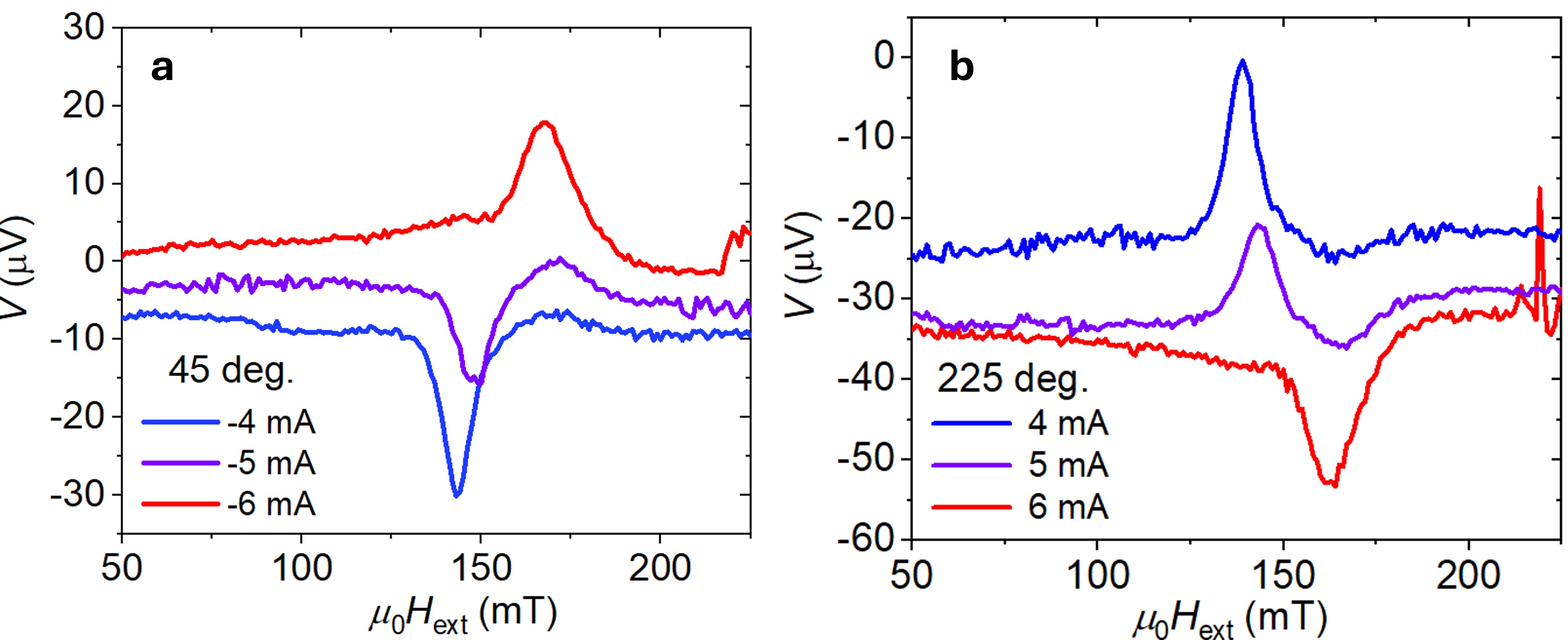}
  \caption{Current-induced magnetic damping control probed by ferromagnetic resonance experiments. Field-swept
FMR voltages measured for \textbf{a} $\phi$ = 45 deg. and \textbf{b} $\phi$ = 225 deg. for 4 GHz while varying $I_\text{dc}$ as specified in the figures. Voltages are offset for each figure. In both cases, clear sign switching can be observed as the magnitude of $I_\text{dc}$ is increased for both polarities for oppositely magnetised cases to satisfy the damping compensation. These are further evidence for the time-averaging magnetisation stabilised around the direction opposite to magnetic field.}
  \label{fig:FMRsignswitch}
\end{figure}

\begin{figure}[t!]
    \centering
    \includegraphics[width=0.5\linewidth]{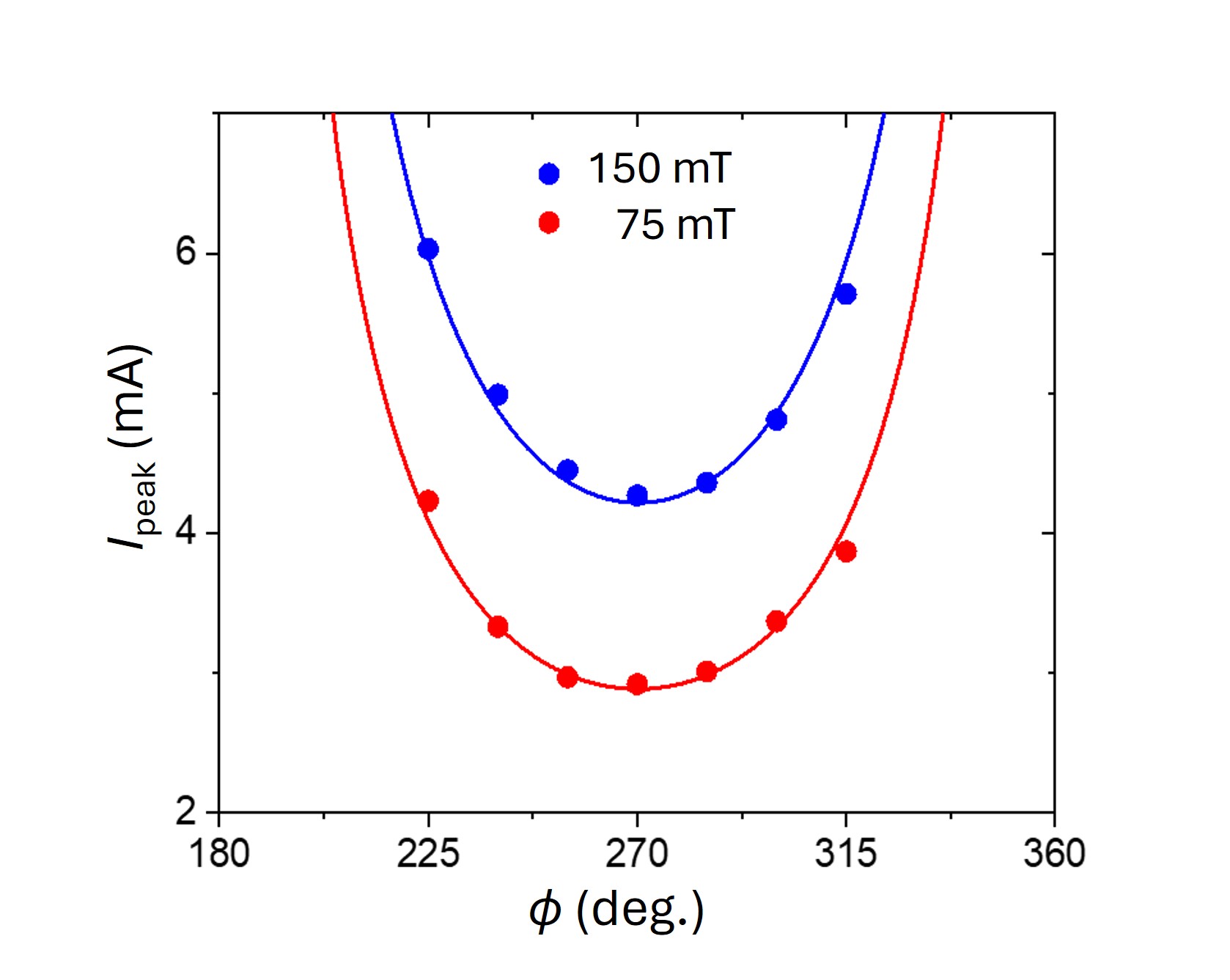}
  \caption{Peak current position ($I_\text{peak}$) as a function of $\phi$ measured at fixed field values (75 and 150~mT) with curves calculated by Eq.(1) in the main text.}
  \label{sfig:mr_peak_angular}
\end{figure}

\end{document}